\documentstyle[12pt]{article}
\advance\textheight by \topskip
\begin{document}
\thispagestyle{empty}
\begin{flushright}
SU--ITP--95--15\\
gr-qc/9508019\\
August 7, 1995\\
\end{flushright}
\vskip 2 cm
\begin{center}
{\LARGE\bf Quantum Cosmology and the Structure}
\vskip 0.3cm
{\LARGE\bf  of Inflationary Universe}\vskip 1.7cm

 {\bf Andrei Linde} \footnote{
Invited talk at the joint Johns Hopkins Workshop -- PASCOS meeting, Baltimore,
March 1995}

\vskip 1.5mm
Department of Physics, Stanford University, \\
Stanford, CA 94305--4060, USA
\end{center}
\vskip 2cm

{\centerline{\large\bf Abstract}}
\begin{quotation}
\vskip -0.4cm
 In this review I will consider several different issues
related to inflation. I will begin with the wave function of the
Universe. This issue is pretty old, but recently there were some new insights
based on
the theory of the self-reproducing inflationary universe. Then we will discuss
stationarity of inflationary universe and the possibility to make predictions
in the
context of quantum cosmology using stochastic approach to inflation. Returning
to   more pragmatic aspects of inflationary theory, we
will discuss inflationary models with $\Omega \not = 1$. Finally, we will
describe several aspects of the  theory of reheating of the Universe based on
the effect of parametric resonance.
\end{quotation}
 \newpage

\section{Wave function of the Universe}

Investigation of the wave function of the Universe goes back to the fundamental
papers by Wheeler and DeWitt \cite{DeWitt}. However,  for a long time it seemed
almost meaningless to apply the notion of the wave function to the Universe
itself,
since the Universe is not a microscopic object. Only with the development of
inflationary cosmology \cite{b14}--\cite{MyBook} it became clear that the whole
Universe could appear from a tiny part of space of the Planck length (at least
in the
chaotic inflation scenario \cite{Chaotic}). Such a tiny piece of space can
appear as a result of quantum fluctuations of metric, which should be studied
in
the context of quantum cosmology.

Unfortunately, quantum cosmology is not a well developed science. This theory
is based on the Wheeler-DeWitt equation, which is the Schr\"{o}dinger equation
for the wave function of the Universe. However,   Bryce DeWitt, one of the
authors of this equation, in some of his talks emphasized that he is not
particularly fond of it. This equation has many solutions, and at the present
time the best method to specify  preferable solutions of this equation (i.e.
its boundary conditions) is based on the Euclidean approach to quantum gravity.
This method  is very powerful, but  some of its applications are not well
justified. In such cases this method may give  incorrect answers,  but  rather
paradoxically sometimes these  answers appear to be correct in application to
some other questions. Therefore it becomes necessary not only to solve the
problem in the Euclidean approach, but also to check, using one's best
judgement, whether the solution is related to the original problem or to
something else.  An alternative approach   is based on the use of stochastic
methods in inflationary cosmology. These methods allows one to understand such
effects as creation of inflationary density perturbations, the theory of
tunneling, and even the theory of self-reproduction of inflationary universe.
Both Euclidean approach and stochastic approach to inflation have their
limitations. However, despite all its problems, quantum cosmology is a very
exciting science which changed dramatically our point of view on the structure
and evolution of the Universe.

We will begin our discussion with the issue of the Universe creation. According
to classical cosmology, the Universe appeared from the singularity in a state
of infinite density. Of course, when the density was greater than the Planck
density $M_{\rm P}^4$ one could not trust classical Einstein equations, but in
many cases there is no demonstrated need to study the Universe creation using
the methods of quantum theory. For example, in the simplest versions of chaotic
inflation scenario \cite{Chaotic} inflation, at the classical level,  could
begin   directly in the initial singularity. However, in certain models, such
as the Starobinsky model \cite{b14}  or the new inflationary universe scenario
\cite{New}, inflation cannot start in a state of infinite density. In such
cases one may speculate about the possibility that inflationary universe
appears due to quantum tunneling ``from nothing.''

The first  idea how one can describe creation of inflationary universe ``from
nothing''
was suggested in 1981 by Zeldovich \cite{Zeld} in  application to the
Starobinsky
model \cite{b14}. His idea was qualitatively correct, but he did not propose
any
quantitative description of this process.  A very important step in this
direction was
made in 1982 by Vilenkin \cite{NothVil}. He suggested to calculate the
Euclidean action on de
Sitter space with the energy density $V(\phi)$,
\begin{equation}
S_E = \int d^4 x \sqrt{-g}\left(-{R\over{16\pi G}}+V(\phi)\right) = -{3M_{\rm
P}^4\over 8  V(\phi)} \ .
\label{Action}
\end{equation}
This action was interpreted as the action on the tunneling trajectory
describing creation of the Universe with the scale factor $a = H^{-1} =
\sqrt{3 M_{\rm P}^2\over 8\pi V}$ from the state with $a = 0$. This would imply
that the probability   of quantum creation of the Universe is given by
\begin{equation}
{\cal P} \propto \exp (-S_E) = \exp \left({3 \over 8  V(\phi)}\right).
\label{Vil1}
\end{equation}
(In  the first three sections of this review we use the system of units with
the Planck mass $M_{\rm P} = 1$.)
A year later this result received a strong support when Hartle and Hawking
reproduced
it by a different though closely related method \cite{HH}. They argued that the
wave function of the ``ground state'' of the Universe with a scale factor $a$
filled
with a scalar field $\phi$ in the semi-classical approximation
is given by
\begin{equation}\label{E31}
\Psi_0(a,\phi)\sim \exp\left(-S_E(a,\phi)\right)\ .
\end{equation}
Here $S_E(a,\phi)$ is the Euclidean action corresponding to
the Euclidean solutions of the Lagrange equation for
$a(\tau)$ and $\phi(\tau)$ with the boundary conditions
$a(0)=a, \phi(0)=\phi$.
The reason for choosing this particular wave function was explained as follows.
Let us consider the Green's function of a particle which moves from
the point $(0,t')$ to the point ${\bf x},t$:
\begin{equation}\label{E32}
<{\bf x},0|0, t'>
 = \sum_n \Psi_n ({\bf x})\Psi_n(0)
\exp\left(iE_{n}(t-t')\right)
 = \int d{\bf x}(t) \exp\left(iS({\bf x}(t))\right)\ ,
\end{equation}
where $\Psi_n$ is a complete set of energy eigenstates
corresponding to the energies $E_n\geq 0$.

To obtain an expression for the ground-state wave function
$\Psi_0({\bf x})$, one should make a rotation
           $t \rightarrow -i\tau$ and take the limit as
$\tau \rightarrow -\infty$\@. In the summation (\ref{E32})
only the term $n=0$ with the lowest eigenvalue $E_0 = 0$
survives, and the
integral transforms into $\int dx(\tau)\exp(-S_E(\tau))$.
Hartle and Hawking have argued that the generalization of
this result to the case of interest in the semiclassical
approximation would yield (\ref{E31}).

The gravitational action corresponding to  one half of  the Euclidean
section $S_4$ of de Sitter space   with $a(\tau) =
H^{-1}(\phi)\cos H\tau$ ($0\leq\tau\leq H^{-1}$) is negative,
\begin{equation}\label{E33}
S_E(a, \phi) = - \frac{3\pi}{4}
    \int d\eta\Bigl[\Bigl(\frac{da}{d\eta}\Bigr)^2 - a^2 +
\frac{8\pi V}{3}a^4\Bigr]
   = - \frac{3}{16 V(\phi)}\ .
\end{equation}
Here $\eta$ is the conformal time, $\eta = \int {d\tau\over a(\tau)}$.
Therefore, according to \cite{HH},
\begin{equation}\label{E34}
\Psi_0(a,\phi)\sim \exp{\Bigl(-S_E(a,\phi)\Bigr)} \sim
\exp\left(\frac{3
}{16V(\phi)}\right) \ .
\end{equation}
By taking a square of this wave function one again obtains eq. (\ref{Vil1}).
The corresponding expression has a very sharp maximum as $V(\phi) \rightarrow
0$\@. This suggests that the probability of finding the Universe in a
state with a large field $\phi$ and having a long stage of
inflation should be strongly suppressed. Some authors consider it as
a strong
argument against the Hartle-Hawking wave function. However, nothing
in the
`derivation' of the Hartle-Hawking wave function tells that it describes
initial
conditions  for inflation. The point of view of  the authors of this wave
function
 was not quite clear. They have written that their wave function gives the
amplitude for the Universe to appear from nothing. On the other hand, Hawking
emphasized \cite{Hawk300} that ``instead of talking about the Universe being
created one should just say: the Universe is.''  This  seems to imply that the
Hartle-Hawking wave function was not designed to describe    initial
conditions at the moment of    the  Universe creation.

Indeed, eq. (\ref{Vil1})  from the very beginning did not seem to apply to the
probability of the Universe creation.   The total energy of matter  in a closed
de
Sitter space with $a(t) =
H^{-1}\cosh  Ht$ is greater than its minimal volume $\sim H^{-3}$ multiplied by
$V(\phi)$, which gives the total energy of the Universe  $E {\
\lower-1.2pt\vbox{\hbox{\rlap{$>$}\lower5pt\vbox{\hbox{$\sim$}}}}\ } M_{\rm
P}^3/\sqrt
V$. Thus the minimal value of the total   energy  of matter contained in a
closed de Sitter universe {\it  grows} when $V$ decreases. For example, in
order to create the
Universe at the Planck density $V \sim M_{\rm P}^4$ one needs no more than the
Planckian energy $M_{\rm P}
\sim 10^{-5}$ g. For the Universe to appear at the GUT energy density  $V
\sim M_X^4$ one needs to create from nothing the Universe with the total energy
of
matter of the order of $M_{\rm Schwarzenegger} \sim 10^2$ kg, which is
obviously
much more difficult. Meanwhile,   eq. (\ref{Vil1}) suggests that it should be
much
easier to create a huge Universe with small $V$ but enormously large total
energy
rather  than a small Universe with large $V$.  This seems very suspicious.

There is one particular place where the derivation (or interpretation) of eq.
(\ref{Vil1}) could go wrong.
The effective  Lagrangian of the scale factor $a$ in (\ref{E33}) has a
wrong sign. The Lagrange equations do not know anything about the sign of the
Lagrangian, so we may simply change the sign before studying the tunneling.
Only
after representing the theory in a conventional form can we  consider tunneling
of
the scale factor. But this then gives us the probability of quantum creation of
the
Universe
\begin{equation}
{\cal P} \propto \exp (-|S_E|) = \exp \left(-{3 \over 8  V(\phi)}\right).
\label{E366}
\end{equation}
This equation predicts that a typical initial value of the
field $\phi$ is given by
$V(\phi)\sim 1$ (if one does not speculate about the
possibility that $V(\phi) \gg 1)$, which leads to a very
long stage of inflation.

Originally I obtained this result by the method described above. However,
because of the ambiguity of the notion of tunneling from the state $a = 0$, I
was not
quite satisfied and decided to look at it from the perspective of derivation of
the
Hartle-Hawking wave function. In this approach  the  problem of the
wrong sign of the Lagrangian appears again, though in a somewhat different
form.
  Indeed,   the
total energy of a closed Universe is zero, being a sum of
the positive energy of matter and the negative energy of the
scale factor $a$.) Thus, the energy $E_n$ of the scale factor is negative, and
in
order to suppress terms with large   negative $E_n$ and to
obtain $\Psi_0$ from
(\ref{E33}) one should rotate $t$ not to $-i\tau$, but to
$+i\tau$. This gives \cite{Creation}
\begin{equation}\label{E35a}
\Psi_0(a,\phi) \sim \exp\Bigl(-|S_E(a,\phi)|\Bigr) \sim\exp
\left(- \frac{3}{16V(\phi)}\right)\ ,
\end{equation}
and
\begin{equation}\label{E36}
P(\phi) \sim|\Psi_0(a,\phi)|^2\sim \exp\Bigl(-2|S_E(a,\phi)|\Bigr)
\sim\exp
\left(- \frac{3}{8V(\phi)}\right) \ .
\end{equation}
 Few months later   this equation was also derived by
 Zeldovich and Starobinsky \cite{ZelStar},
 Rubakov \cite{Rubakov}, and Vilenkin \cite{Vilenkin} using the methods similar
to the first
method mentioned above (investigation of   tunneling in the theory with the
wrong
sign of the Lagrangian). The corresponding wave function (\ref{E35a}) was
called ``the tunneling wave function.''

An obvious objection against this result is that it may be incorrect
 to use  different  ways of rotating $t$ for quantization of the
scale
factor and of the  scalar field. However, the idea that a consistent
quantization
of
matter coupled to gravity can be accomplished  by a proper choice
of a complex contour of integration may be too optimistic.  We know, for
example,
that despite
many attempts to suggest  Euclidean formulation for nonequilibrium quantum
statistics
or for the field theory in a nonstationary background,   such
formulation does not exist.
It is quite clear from (\ref{E32}) that the $t \rightarrow -i\tau$
trick
would not work if the spectrum $E_n$ were not bounded from below.
Absence of equilibrium, of any simple
stationary ground state seems to be  a typical situation in quantum
cosmology. In
some cases where a stationary or quasistationary ground state does
exist, eq. (\ref{Vil1}) may be correct, see the next Section.
In a more general situation
it may be very difficult to obtain any simple expression for the wave
function of the Universe. However, in certain limiting cases this
problem is relatively simple. For example, at present the scale
factor
$a$ is very big and it changes very slowly, so one can consider it
to be a C-number and quantize matter fields only by rotating $t
\rightarrow
-i\tau$\@. On the other hand, in the
inflationary universe the evolution of the scalar field is  very
slow;
during the typical time intervals $O(H^{-1})$ it behaves essentially
as
a  classical field, so one can describe the process of the creation
of
an inflationary universe  filled with a homogeneous scalar field by
quantization of
the
scale factor $a$ only and by rotation $t \rightarrow i\tau$.

Derivation of equations (\ref{Vil1}), (\ref{E36}) and their interpretation is
far
from being rigorous, and therefore even now it remains the subject of debate.
In
our opinion, the Hartle-Hawking wave function describes not the Universe
creation,
but the fluctuations of the Universe near its de Sitter ground state, under the
condition that such a state  exists, see next section. Meanwhile the
distribution
(\ref{E36})  is related to the probability of creation of inflationary universe
from
nothing (or from the space-time foam). However,  the two different derivations
of
this probability distribution emphasize two slightly different features of the
process. Investigation of tunneling should give  the probability of   quantum
creation of the Universe of a size $H^{-1}$ from the Universe with $a = 0$.
Meanwhile wave function of the ``ground state'' should  give  information
about some kind of probability distribution of various Universes in the
space-time
foam.   We will not concentrate here on this subtle difference since we
believe that it would bring us far away from the domain of applicability of our
approach. Also, we should emphasize again that quantum tunneling is necessary
only if one cannot use the classical trajectory. In the  Starobinsky model
\cite{b14}, as well as in  the new inflationary universe scenario \cite{New},
creation of the Universe ``from nothing'' appears to be one of the most natural
mechanisms for inflation to begin. Meanwhile, in the simplest version of
chaotic inflation scenario the process of inflation formally may begin directly
in the singularity, in a state with infinitely large $V(\phi)$, without any
need for quantum tunneling. However, quantum tunneling in that case is possible
as well, since for $V(\phi) \sim 1$ the probability of quantum creation of
inflationary universe is not exponentially suppressed.

In the next section we will discuss stochastic approach to quantum
cosmology. Within this approach equations  (\ref{Vil1}) and (\ref{E36}) can be
derived in a much more clear and rigorous way, but they will have somewhat
different interpretation.

\

\section{Wave function of the Universe and stochastic approach to inflation}

 The first models of inflation were  based on the standard assumption
of the
big bang theory that the Universe was created at a single moment of
time in a
state with the Planck density, and that it was hot and large (much
larger than
the Planck scale $M_{\rm P}^{-1} =1$) from the very beginning. The  success
of
inflation in solving internal problems of the big bang  theory
apparently
removed the last doubts concerning the  big bang cosmology.  Even in our
quantum mechanical treatment of the Universe production we still used the
standard idea that the Universe as a whole can be described by one scale factor
$a$, and its creation should be considered as a process beginning from $a = 0$.
Meanwhile during the last ten years the inflationary theory  has
broken
the umbilical cord connecting it with the old big bang theory, and
acquired an
independent life of its own. For the practical purposes of
description of the
observable part of our Universe one may still speak about the big bang.
However, if one tries to understand the beginning of the Universe, or
its end,
or its global structure, then some of the notions of the big bang
theory become
 inadequate.

For example,   in the  chaotic inflation
scenario \cite{Chaotic} even without taking into account quantum effects there
was no need to assume that the whole Universe
appeared from
nothing at a single moment of time associated with the big bang, that
the
Universe was hot from the very beginning and that the inflaton scalar
field
$\phi$ which drives inflation originally occupied the minimum of its
potential
energy. On the other hand, it was found that if the Universe in this scenario
contains at least one
inflationary domain of a size of horizon  (`$h$-region') with a
sufficiently
large and homogeneous scalar field $\phi$, then this domain will
permanently
produce new $h$-regions of a similar type due to quantum fluctuations
\cite{b19,b20}. This process  occurs   in  the old, new  and extended inflation
scenario as well \cite{b51}--\cite{EtExInf}.  Thus,  instead
of a single
big bang producing a one-bubble Universe, we are speaking now about
inflationary bubbles producing new bubbles, producing new bubbles,
 {\it  ad
infinitum}. In this sense, inflation is not a short intermediate
stage of
duration $\sim 10^{-35}$ seconds, but a self-regenerating process,
which occurs
in some parts of the Universe even now, and which will continue
without end.

It is extremely complicated to describe an inhomogeneous self-reproducing
Universe.  Fortunately, there
is a
particular
kind of stationarity of the process of the Universe self-reproduction
which
makes things  more regular. Due to the no-hair theorem for de Sitter
space, the
process of production of new inflationary domains occurs
independently of any
processes outside the horizon. This process depends only on the
values of the
fields inside each $h$-region   of  radius $H^{-1}$. Each time  a new
inflationary $h$-region is created during the Universe expansion, the
physical
processes inside this region  will depend only on the properties of
the fields
inside it, but  not  on the `cosmic time'  at which it was created.

In addition to this most profound stationarity,  which we will call  {\it
local
stationarity}, there may also exist some
simple stationary probability distributions which may allow us to
say, for
example,  what  the probability is of finding a given field $\phi$ at
a given
point.  To examine this possibility one should consider the
probability
distribution $P_c(\phi, \chi,t)$, which  describes the probability
of finding
the field $\phi$ at a given point at a time $t$, under the condition
that at
the time $t=0$ the field $\phi$ at this point was equal to $\chi$. The same
function   also describes the probability
that the
scalar field which at  time $t$ was equal to $\phi$, at some earlier
time $t=0$
was equal to $\chi$.

  The  probability distribution $P_c$  is  in fact the
probability
distribution per  unit volume in  {\it  comoving coordinates} (hence
the index
$c$ in  $P_c$), which do  not change during expansion of the
Universe.  By
considering this probability distribution   we neglect the main
source of the
self-reproduction of inflationary domains, which is the exponential
growth of
their volume. Therefore, in addition to   $P_c$, we introduced the
probability
distribution $P_p(\phi,\chi,t)$, which describes the probability to
find a
given field configuration in a unit physical volume \cite{b19}. In
the situations where one of these distributions can be stationary,  we will
speak
about  {\it  global stationarity}.

Let us remember some details of stochastic approach
to
inflation.
Consider the simplest model of chaotic inflation based on the
theory of
a  scalar field $\phi$ minimally coupled to gravity, with the
effective
potential $V(\phi)$. If the classical field $\phi$ is sufficiently
homogeneous
in some domain of the Universe, then its behavior inside this domain
is
governed by the equation $3H\dot\phi = -dV/d\phi$, where $H^2 =
\frac{8\pi V(\phi)}{3 }$.
Investigation of these equations has shown that  in all power-law potentials
$V(\phi)\sim \phi^n$ inflation occurs
at $\phi > \phi_e \sim n/6$. In the theory with an exponential potential
$V(\phi)\sim
e^{\alpha \phi}$ inflation ends only if we bend down the potential at some
point $\phi_e$;
for definiteness we will take $\phi_e = 0$ in this theory.

Inflation stretches all initial inhomogeneities. Therefore, if the
evolution of the Universe were governed solely by classical equations
of
motion, we would end up with an extremely smooth Universe with no
primordial
fluctuations to initiate the growth of galaxies.
Fortunately, new density perturbations are generated during inflation
due to
quantum effects. The wavelengths of all  vacuum
fluctuations of the scalar field $\phi$ grow exponentially in the
expanding
Universe. When the wavelength of any particular fluctuation becomes
greater
than $H^{-1}$, this fluctuation stops oscillating, and its amplitude
freezes at
some nonzero value $\delta\phi (x)$ because of the large friction
term
$3H\dot{\phi}$ in the equation of motion of the field $\phi$\@. The
amplitude
of this fluctuation then remains
almost unchanged for a very long time, whereas its wavelength grows
exponentially. Therefore, the appearance of such a frozen fluctuation
is
equivalent to the appearance of a classical field $\delta\phi (x)$
that does
not vanish after averaging over macroscopic intervals of space and
time.

Because the vacuum contains fluctuations of all wavelengths,
inflation leads to
the creation of more and more perturbations of the classical field
with
wavelengths greater than $H^{-1}$\@. The average amplitude of such
perturbations generated during a time interval $H^{-1}$ (in which the
Universe
expands by a factor of e) is given by
\begin{equation}\label{E23}
|\delta\phi(x)| \approx \frac{H}{2\pi}\ .
\end{equation}
The phases of each wave are random. Therefore,  the sum of all waves
at a given
point fluctuates and experiences Brownian jumps in all directions in
the space
of fields.

One can describe  the stochastic behavior of the inflaton
field using diffusion equations for the  probability
distribution $P_c(\phi,t|\chi)$. The first equation is called the backward
Kolmogorov equation,
\begin{equation} \label{Starb}
\frac{\partial  P_c(\phi,t|\chi)}{\partial t} =
  \frac{H^{3/2}(\chi)}{8\pi^2}\, \frac{\partial
}{\partial\chi}
 \left({H^{3/2}(\chi)}  \, \frac{\partial }{\partial\chi}
P_c(\phi,t|\chi)
\right)
 -  \frac{V'(\chi)}{3H(\chi)} \frac{\partial }{\partial\chi}
P_c(\phi,t|\chi)   \ .
\end{equation}
In this equation one considers the value of the field $\phi$ at the
time  $t$
as a constant,  and finds the time dependence of the probability that
this
value  was reached  during the time  $t$ as a result of
diffusion of the
scalar field from different possible initial values $\chi \equiv
\phi(0)$.

The second equation is the adjoint to the first one; it is called the
forward
Kolmogorov equation, or the Fokker-Planck equation \cite{Star},
\begin{equation}\label{E3711}
\frac{\partial P_c(\phi,t|\chi)}{\partial t} =
   \frac{\partial }{\partial\phi}
 \left( \frac{H^{3/2}(\phi)}{8\pi^2}\, \frac{\partial }{\partial\phi}
\Bigl(
{H^{3/2}(\phi)}  P_c(\phi,t|\chi) \Bigr) +
\frac{V'(\phi)}{3H(\phi)} \, P_c(\phi,t|\chi)\right) \ .
\end{equation}

One may try to find a stationary solution of equations (\ref{Starb}),
(\ref{E3711}), assuming that  $\frac{\partial
P_c(\phi,t|\chi)}{\partial t} =
0$.
The simplest stationary solution  (subexponential factors being
omitted) would
be \cite{Star,Mijic,LLM}
\begin{equation}\label{E38a} P_c(\phi,t|\chi) \sim
\exp\left({3\over 8 V(\phi)}\right)\cdot \exp\left(-{3\over 8
V(\chi)}\right) \ .
\end{equation}
This looks like a miracle: The first term  in
this expression is equal to the square of the Hartle-Hawking wave function of
the
Universe  (\ref{Vil1}), whereas the second one gives the square of the
tunneling
wave
function (\ref{E36})! And we obtained this result without any   ambiguous
considerations based on Euclidean approach to quantum cosmology!

At  first glance, this result gives a direct confirmation and a
simple physical
interpretation of  both the Hartle-Hawking  wave function of the
Universe  {\it
and} the tunneling wave function. First of all, we see that the distribution of
the probability to find the Universe in a state with the field $\phi$ is
proportional to $\exp\left({3\over 8 V(\phi)}\right)$. Note that we are
speaking here about the state of the Universe rather than the probability of
its creation. Meanwhile, the probability that the Universe emerged from the
state with the field $\chi$ is proportional to $\exp\left(-{3\over 8
V(\chi)}\right)$. Now we are speaking about the probability that a given part
of the Universe was created from the state with the field $\chi$,   and the
result coincides with our result for the probability of the quantum creation of
the Universe, eq.  (\ref{E36}).

This would be a great peaceful resolution of the conflict between the two wave
functions.
However, the situation is much more complicated. In all realistic
cosmological
theories, in which $V(\phi)=0$ at its minimum, the Hartle-Hawking
distribution
$\exp\left({3\over 8 V(\phi)}\right)$ is not normalizable. The source
of this
difficulty can be easily
understood: any stationary distribution may exist only due
to  compensation of the classical flow of the field $\phi$
downwards to the minimum of $V(\phi)$ by the diffusion motion
upwards. However, diffusion of the field $\phi$ discussed
above exists only during inflation. Thus, there is no diffusion
motion upwards
from the region $\phi < \phi_e$. Therefore expression (\ref{E38a}) is not a
true solution of  equation
(\ref{E3711});  all solutions  with the proper  boundary conditions at $\phi =
\phi_e$
(i.e. at
the end of inflation)  are non-stationary (decaying) \cite{b19}.

It is possible to use the solution (\ref{E38a}) in the cases where   the state
can be  quasistationary. For example, in  the case when the effective potential
has a local minimum with a sufficiently large $V$, this distribution   gives a
correct expression for the probability of the Hawking-Moss tunneling
\cite{Star}. We were unable to find a situation in the context of inflationary
cosmology where one could ascribe a more fundamental meaning to the
Hartle-Hawking wave function, but of course this might be a result of our own
limitations.

One can get an additional insight by investigation of the probability
distribution $P_p$.  In order to do so, one should write stochastic equations
for ${\cal V}(\phi,t|\chi)$, where ${\cal V}(\phi,t|\chi)$ is the total volume
of domains with the field $\phi$ originated from the domains containing field
$\chi$. The system of
stochastic
equations for ${\cal V}(\phi,t|\chi)$  can be obtained from eqs.
(\ref{Starb}),
(\ref{E3711}) by   adding the term $3H{\cal V}$, which appears due to the
growth of
physical volume of the Universe by the factor $1 + 3H(\phi)\, dt$
during each
time interval $dt$ \cite{Nambu}--\cite{LLM}:
\begin{equation} \label{Starbx}
\frac{\partial  {\cal V}}{\partial t} =
  \frac{H^{3/2}(\chi)}{8\pi^2}\, \frac{\partial
}{\partial\chi}
 \left( {H^{3/2}(\chi)}  \, \frac{\partial
{\cal V}}{\partial\chi}
\right)
 -  \frac{V'(\chi)}{3H(\chi)} \frac{\partial  {\cal V}}{\partial\chi}
  +3H(\chi)  {\cal V} \ ,
\end{equation}
\begin{equation}\label{E372}
\frac{\partial {\cal V}}{\partial t} =
    \frac{\partial }{\partial\phi}
 \left( \frac{H^{3/2}(\phi)}{8\pi^2}\, \frac{\partial }{\partial\phi}
\Bigl(
  {H^{3/2}(\phi)}   {\cal V} \Bigr) +
\frac{V'(\phi)}{3H(\phi)} \, {\cal V}\right)  +  3H(\phi)  {\cal V}\ .
\end{equation}
To find solutions of these equations one must specify boundary
conditions. Behavior of solutions of  these equations typically is not very
sensitive to the
boundary conditions at the boundary  $\phi = \phi_e$ where inflation ends; it
is sufficient to assume that the diffusion
coefficient (and, correspondingly, the first terms in the r.h.s. of equations
(\ref{Starbx}), (\ref{E372})) vanish for $\phi < \phi_e$.
The
conditions
at the Planck boundary $\phi = \phi_p$  play a more important role.
In what follows we will  assume that inflation ceases to exist at $\phi >
\phi_p$ \cite{LLM}. This
leads to the boundary condition
\begin{equation}\label{PlanckBound}
 {\cal V}(\phi_p,t|\chi)= {\cal V}(\phi,t|\chi_p) = 0 \ ,
\end{equation}
where $V(\phi_p) \equiv V(\chi_p) = O(1)$.

One may try to obtain solutions of equations (\ref{Starbx}),
(\ref{E372}) in the
form of the following series of system of
eigenfunctions of the
pair of adjoint linear operators defined by the left hand sides of
the
equations below:
\begin{equation} \label{eq14}
{\cal V}(\phi,t|\chi) =
\sum_{s=1}^{\infty} { e^{\lambda_s t}\, \psi_s(\chi)\, \pi_s(\phi) }
\ .
\end{equation}
Indeed, this gives us a solution of eq. (\ref{E372}) if
\begin{equation} \label{eq15}
   \frac{H^{3/2}}{8\pi^2} \frac{\partial
}{\partial\chi}
 \left({H^{3/2}}  \frac{\partial }{\partial\chi}
\psi_s(\chi) \right)
 - \frac{V'}{3H} \frac{\partial }{\partial\chi}
\psi_s(\chi)
 + 3H  \cdot \psi_s(\chi) =
\lambda_s \, \psi_s(\chi)  \ ,
\end{equation}
and
\begin{equation} \label{eq17}
   \frac{\partial }{\partial\phi}
 \left( \frac{H^{3/2}}{8\pi^2} \frac{\partial }{\partial\phi}
\left({H^{3/2}}  \pi_j(\phi) \right) \right)
 + \frac{\partial }{\partial\phi} \left( \frac{V'}{3H} \,
\pi_j(\phi) \right)
 + 3H\cdot \pi_j(\phi) =
\lambda_j \, \pi_j(\phi)  \ .
\end{equation}

In our case (with regular boundary conditions) one can easily show
that the
spectrum of $\lambda_j$ is discrete and bounded from above. Therefore
the
asymptotic solution for ${\cal V}(\phi,t|\chi)$ in the limit $t
\rightarrow \infty$ is given by
\begin{equation} \label{eq22}
{\cal V}(\phi,t|\chi) = e^{\lambda_1 t}\, \psi_1(\chi) \,
\pi_1(\phi)\, \cdot \left(1 + O\left( e^{-\left(\lambda_1 - \lambda_2
\right) t} \right) \right) \ .
\end{equation}
Here $\psi_1(\chi)$ is the only positive eigenfunction of eq.
(\ref{eq15}),
$\lambda_1$ is the corresponding (real) eigenvalue, and $\pi_1(\phi)$
 is the eigenfunction
of the conjugate operator (\ref{eq17}) with the
same eigenvalue $\lambda_1$\@. Note, that $\lambda_1$ is
the highest eigenvalue, $\mbox{Re} \left( \lambda_1 - \lambda_2
\right) > 0 $\@.    This means that the distribution
\begin{equation} \label{eq22aa}
{P}_p(\phi,t|\chi) = e^{-\lambda_1 t}
\,{\cal V}(\phi,t|\chi)
\end{equation}
gradually converges to the time-independent normalized distribution
\begin{equation} \label{eq22a}
{P}_p(\phi,\chi) \equiv
{P}_p(\phi,t \rightarrow \infty|\chi) =  \psi_1(\chi) \,
\pi_1(\phi) \
{}.
\end{equation}
It is this stationary distribution that we were looking for.    $
{P}_p(\phi,\chi)$ gives   the
fraction of
the volume of the Universe occupied by the  field $\phi$, under the
condition
that the corresponding part of the Universe at some time in the past
contained
the field $\chi$.  The
remaining problem is to find the functions $\psi_1(\chi)$ and
$\pi_1(\phi)$,
and to check that all assumptions about the boundary conditions which
we made on the way to eq. (\ref{eq22}) are actually satisfied.

We have solved this problem for chaotic inflation in a wide class of
theories
including the theories with polynomial and exponential effective
potentials
$V(\phi)$ and found the corresponding stationary distributions \cite{LLM}. Here
we will present some of our results for the
theory
${\lambda\over 4}\phi^4$.

 Solution of equations (\ref{eq15}) and (\ref{eq17}) for
$\psi_1(\chi)$ and
$\pi_1(\phi)$   shows that
these
functions are extremely small at  $\phi\sim \phi_e$ and $\chi\sim
\chi_e$, where $\phi_e\sim \chi_e \sim 1$ correspond to the end of inflation.
These functions
grow at large $\phi$ and $\chi$,  then rapidly decrease, and vanish
at $\phi =
\chi=  \phi_p$. With a decrease of $\lambda$ the solutions become
more and more
sharply peaked near the Planck boundary. To give a physical interpretation to
our solutions, it will be convenient   to parametrize $\lambda_1$ in the
following form: $ \lambda_1= d(\lambda) H_{\rm \max}(\lambda)$. Here $d$ is the
so-called fractal dimension of inflationary universe \cite{ArVil,LLM}, and
$H_{\rm \max} $ is the maximal value of the Hubble constant in the model under
consideration. For example, in the models where inflation ceases to exist at
the Planck density $V(\phi) = 1$ the maximal value of the Hubble constant is
given by $2\sqrt{2\pi\over 3}$.  The eigenvalues $d(\lambda)$
corresponding to
different coupling constants $\lambda$ are given by the following
table:
\begin{center}
\begin{tabular}{|c|c|c|c|c|c|c|c|}
\hline \hline
$\lambda$ & $1$ & $10^{-1}$ & $10^{-2}$ & $10^{-3}$ & $10^{-4}$
& $10^{-5}$ & $10^{-6}$ \\
\hline
 $d$ & 0.9719 & 1.526 & 1.915 & 2.213 & 2.438 & 2.604 & 2.724
\\
\hline \hline
\end{tabular}
\end{center}
As we see, in the limit $\lambda \to 0$ the fractal dimension $d(\lambda)$
grows   toward
the usual space dimension $3$.

It is very interesting to study the behavior of $P_p$ at small $\phi$ and
$\chi$, i.e. at the stage which determines the structure of the observable part
of the Universe. One could expect to find a dependence similar to the one given
by eq. (\ref{E38a}), i.e. $P_p\sim \exp\left({3\over 8 V(\phi)}\right)\cdot
\exp\left(-{3\over 8
V(\chi)}\right)$. Indeed, this remains true for the dependence of $P_p$ on
$\chi$. Meanwhile, since there is no diffusion term at $\phi < \phi_e$, the
solution at small $\phi> \phi_e$ should match the solution obtained by
neglecting the first (diffusion) term at $\phi < \phi_e$. As a result, instead
of the product of the Hartle-Hawking and the tunneling solution for the theory
${\lambda\over 4} \phi^4$ for  small $\phi$ and $\chi$ (for $\phi, \chi <
\lambda^{-1/8}$)  we obtain
 \begin{equation}\label{SMALLPHI}
{\cal V}(\phi,\chi,t)\, =\, e^{ d(\lambda)  H_{\rm \max}t}~ {P}_p(\phi,\chi)\,
\sim\,  e^{ d(\lambda)  H_{\rm \max}t}~  \phi^{\sqrt{6\pi\over
\lambda}\lambda_1}~ \exp\left(-{3\over 8
V(\chi)}\right)      \ .
\end{equation}
Thus, the square of the tunneling wave function is here, but the square of the
Hartle-Hawking wave function dropped away.  The dependence of  ${\cal
V}(\phi,\chi,t)$ on $\chi$ and $\phi$ is extremely sharp.   For example, for
the realistic value
$\lambda \sim 10^{-13}$
one has $ {P}_p(\phi,\chi) \sim  e^{10^{13}\chi^{-4}}  \, \phi^{10^8}$.

The factor $e^{ d(\lambda)  H_{\rm \max}t}$ controls the speed of exponential
expansion of
the volume filled by a given field $\phi$.  {\it  This speed does not
depend on
the field $\phi$}, and has the same order of magnitude as the speed
of
expansion at the Planck density.  One should emphasize    that the factor $e^{
d(\lambda)  H_{\rm \max}t}$
  gives the rate of growth of the combined volume
of all domains with a given field $\phi$ (or of all domains
containing matter with a given density)  {\it   not only at very
large $\phi$, where quantum fluctuations are large, but at small
$\phi$ as well, and even after inflation} \cite{LLM}. This
result may seem absolutely unexpected, since the volume of each
particular inflationary domain grows like $e^{3H(\phi)t}$, and
after inflation the law of expansion becomes completely
different.  One should distinguish, however, between the growth
of each particular domain, accompanied by a decrease of density
inside it, and the growth of the total volume of all domains
containing matter with a given (constant) density. In the
standard big bang theory the second possibility did not exist,
since the energy density was assumed to be  the same in all
parts of the Universe (``cosmological principle''), and it was
not constant in time.

The reason why there is a universal expansion rate $e^{\lambda_1 t}$
can be understood as follows. Because of the self-reproduction
of the Universe there always exist many domains with $\phi \sim
\phi_{p}$,  and their combined volume grows almost as fast as
$e^{3H(\phi_p) t}$. Then the field $\phi$ inside some of these
domains decreases. The  total volume of domains containing some
small field $\phi$ grows not only due to expansion $\sim
e^{3H(\phi_p) t}$, but mainly due to the unceasing process of
expansion of domains with large $\phi$ and their subsequent
rolling (or diffusion) towards small $\phi$.

The distribution ${P}_p(\phi, \chi)  =  \psi_1(\chi) \,
\pi_1(\phi)$
which we have obtained  does not depend on time $t$. However, in
general
relativity one may use many different time parametrizations, and the
same
physics can be described
differently in different `times'. One of the most natural choices of
time in
the context of stochastic approach to inflation is the time $\tau  =
\ln{{a\left(x, t \right) \over a(x,0)}} =  \int{H(\phi(x,t),t)\ dt}$
\cite{Star,Bond}.  Here $a\left(x, t \right)$ is a local value of the
scale
factor in the inflationary universe. By using this time variable, we
were able
to obtain not only numerical solutions to the stochastic equations,
but also
simple asymptotic expressions describing these solutions. For
example, for the
theory ${\lambda\over 4 } \phi^4$ both the eigenvalue $\lambda_1$ and
the `fractal dimension' $d_f$ (which in this case refers both to the
Planck
boundary at $\phi_p$ and to the end of inflation at $\phi_e$) are
given by $d_f
= \lambda_1 \sim 3-1.1\, \sqrt \lambda$, and the stationary
distribution is \cite{LLM}
\begin{eqnarray}
{P}_p(\phi,\chi) &\sim &  \exp\Bigl(-{3\over
8V(\chi)}\Bigr)\,
\Bigl({1\over V(\chi)+0.4} - {1\over1.4}\Bigr)\, \cdot \,   \phi
\,\exp\Bigl(-{\pi\, (3-\lambda_1)\phi^2}\Bigr) \nonumber \\
 &\sim & \exp\Bigl(-{3\over 2 \lambda\, \chi^4}\Bigr)\, \Bigl({4\over
\lambda
\chi^4+1.6} - {1\over1.4}\Bigr)\, \cdot \,   \phi \, \exp\Bigl(- 3.5
\sqrt\lambda\phi^2\Bigr)\ .
\end{eqnarray}
 The first factor again coincides with the square of the tunneling
wave
function, and again there is no trace of the Hartle-Hawking wave function. This
expression is valid  in the whole
interval from
$\phi_e$ to  $\phi_p$ and it correctly describes asymptotic behavior
of
${P}_p(\phi,\chi)$ both at  $\chi \sim \chi_e$ and at
$\chi \sim
\chi_p$.

A similar investigation can be carried out for the theory $V(\phi) =
V_o\
e^{\alpha\phi}$. The corresponding solution is
\begin{eqnarray}
 {P}_p(\phi,\chi) &\sim &  \exp\Bigl(-{3\over
8V(\chi)}\Bigr)\,
\Bigl({1\over V(\chi)} - {1}\Bigr)\, \cdot \,  \Bigl({1\over V(\phi)}
-
{1}\Bigr)\, V^{-1/2}(\phi)\ .
\end{eqnarray}
This expression gives a  rather good approximation for
${P}_p(\phi,\chi)$ for all $\phi$ and  $\chi$.

The main result  is that under certain conditions the
properties of
our   Universe  can be described by a time-independent probability
distribution, which we have found for theories with polynomial and
exponential
effective potentials. Thus,  inflation  solves many problems of the big bang
theory and
ensures that
this theory provides an excellent description of the local structure
of the
Universe. However,  after making all kinds of improvements of  this
theory, we
are now winding up with a model of a stationary Universe, in
which the
notion of the big bang    loses its dominant position, being removed
to the indefinite past.

\section{\label{predictions} Predictions in quantum cosmology}
\subsection{Moderate approach: comparing probabilities within the same
Universe}

When inflationary theory was first formulated, we did not know
how much  it was going to influence our  understanding of the structure of the
Universe.  We were happy   that inflation provided an easy explanation of the
homogeneity of the Universe. However, we did not know that the same theory
simultaneously predicts that on the extremely large scale the Universe becomes
entirely inhomogeneous, and that this inhomogeneity is good, since it is one of
the manifestations of the process of self-reproduction of inflationary
Universe.

The new picture of the Universe which emerges now is very unusual, and we are
still in the process of learning how to ask proper questions in the context of
the new cosmological paradigm. Previously we assumed  that we live in a
Universe which has the same properties everywhere (``cosmological principle'').
 Then one could make a guess about the most natural initial conditions in the
Universe, and all the rest followed almost automatically. Now we learned that
even if one begins with a non-uniform Universe, later it   becomes extremely
homogeneous on a very large scale. However, simultaneously it becomes
absolutely non-uniform on a much greater scale. the Universe becomes divided
into different exponentially large regions where the laws of low-energy physics
can be different. In certain cases the relative fraction of volume of the
Universe in a state with given fields or with a given density does not depend
on time, whereas the total volume of all parts of the Universe continues
growing exponentially.

This change of  the picture of the world is important by itself. However, it
would be even better if we could use it to make certain predictions based on
this picture. In this situation the   problem of introduction of a proper
measure of probability becomes
most important.  One of the most natural choices of such measure is given by
the probability distribution $ P_p(\phi,\chi,t)$. The hypothesis behind this
proposal is that we are typical, and therefore we live in those parts of the
Universe where   most other people do.  The total number of people which can
live in domains with given properties should be proportional to the total
volume of these domains. There are two versions of this hypothesis, the
moderate and the radical ones. The moderate version is based on investigation
of $P_p(\phi,\chi,t)$ \cite{b19,LLM,GBLL}. If this distribution  is stationary,
then it seems
reasonable to use it as a measure of the total volume of   domains with any
particular properties at any
given moment of time $t$.

The first example of this approach is given by the consideration of the axion
problem. In the non-inflationary cosmology it was shown that the axion mass
should be greater than $10^{-5}$ eV in order to avoid having too much energy
stored in the axion field \cite{axion}.
However, the derivation of this constraint fails in inflationary cosmology if
one takes into account quantum fluctuations of the axion field and eternal
production of domains where this field takes all its possible values. Then it
can be shown  that life of our type can appear only in those domains where the
axion field is sufficiently small and under certain conditions discussed in
\cite{LinAx} the standard constraint $m_a >10^{-5}$ eV disappears.

 Another interesting example is given by  the probability distribution for
finding the most probable
values of the effective gravitational constant in the Brans-Dicke inflationary
cosmology \cite{GBLL}. We have shown there that inflation in
the Brans-Dicke theory leads to division of the Universe into different
exponentially large domains with different values of the gravitational
constant, and, correspondingly, with different values of density perturbations.
Then one can use  the probability distribution $P_p(\phi,\chi,t)$ to find most
probable values of the gravitational constant.  In this approach it is possible
either to explain the anomalously large value of the Planck mass, or at least
relate it to certain small parameters in the theory, e.g. to the small
anisotropy of the microwave background radiation. Note, that the very language
which we are using may sound somewhat strange. Indeed, typically the purpose is
to express the anisotropy of the microwave background radiation via some
fundamental parameters of the theory. In our case the Planck mass is not
fundamental, and its  value is anomalously large in those domains where the
microwave background radiation is anomalously small.

In what follows
I will briefly describe some    nonperturbative effects  which may lead to a
considerable local deviation of density from the critical density of a flat
Universe \cite{OPEN}.

Let us consider all parts of inflationary universe which contain a given field
$\phi$ at a given moment of time $t$. One may wonder, what was the value of
this field in those domains at the moment $t - H^{-1}$ ? The  answer is simple:
One should add to $\phi$ the value of its classical drift $\Delta \phi$ during
the time $H^{-1}$,  $\Delta \phi = \dot\phi H^{-1}$. One should also add the
amplitude of a quantum jump $\delta \phi$. The typical   jump  is given by
$\delta \phi = \pm {H\over 2\pi}$. At the last stages  of inflation this
quantity is by many orders of magnitude smaller than $\Delta \phi$. However, in
which sense   jumps $\pm {H\over 2\pi}$ are typical? If we consider any
particular initial value of the field $\phi$, then the typical jump from this
point is indeed given by $\pm {H\over 2\pi}$. However, if we are considering
all domains with a given $\phi$ and trying to find all those domains from which
the field $\phi$ could originate back in time, the answer may be quite
different. Indeed,   the total volume of all domains with a given field $\phi$
at any moment of time $t$ strongly depends on $\phi$:   \,${P}_p(\phi)
\sim   \phi^{\sqrt{6\pi\over \lambda}\lambda_1} \sim \phi^{10^8}$, see eq.
(\ref{SMALLPHI}). This means that the total volume of all domains which could
jump towards the given field $\phi$ from the value $\phi +\delta \phi$ will be
enhanced by a large  additional factor $ {{P}_p(\phi +\delta \phi)\over
{P}_p(\phi)} \sim   \Bigl(1+{\delta\phi\over \phi}\Bigr)^{\sqrt{6\pi\over
\lambda}\lambda_1}$. On the other hand, the probability of large jumps
$\delta\phi$ is suppressed by the Gaussian factor
$\exp\Bigl(-{2\pi^2\delta\phi^2\over H^2}\Bigr)$.
One can easily verify that the product of these two factors has a sharp maximum
at $\delta\phi = \lambda_1 \phi   \cdot {H\over 2\pi}$, and the width of this
maximum is of the order ${H\over 2\pi}$. In other words, most of the domains of
a given field $\phi$ are formed due to   jumps which are greater than the
``typical'' ones by a factor $\lambda_1 \phi \pm O(1) $.

Our part of the Universe in the inflationary scenario with $V(\phi) =
{\lambda\over 4} \phi^4$ is formed at $\phi~\sim~5$ (in the units $M_{\rm P} =
1$),
and the constant $\lambda_1 \approx   2\sqrt{6\pi} \sim 8.68$ for our choice of
boundary conditions. This means that our part of the  Universe should be
created as a result of a jump down which is about  $\lambda_1 \phi \sim 40$
times greater than the standard jump. The standard jumps   lead to   density
perturbations of the   amplitude $\delta\rho \sim 5\cdot10^{-5} \rho_c$ (in the
normalization of  \cite{MyBook}). Thus, according to our nonperturbative
analysis, we should live inside a region where density is smaller than the
critical density by about $\delta\rho \sim 2\cdot10^{-3} \rho_c$. As we already
mentioned, the probability of such fluctuations should be suppressed  by
$\exp\Bigl(-{2\pi^2\delta\phi^2\over H^2}\Bigr)$, which in our case gives the
suppression factor $ \sim  \exp(-10^3)$. It is well known that exponentially
suppressed perturbations typically give rise to spherically symmetric bubbles.
Note also, that the Gaussian distribution suppressing the amplitude of the
perturbations refers to the amplitude of a perturbation in its maximum. It is
possible that we live not in the place corresponding to the maximum of the
fluctuation. However, this could only happen   if the nonperturbative jump down
was even greater in the amplitude that we expected. Meanwhile, as we already
mentioned, the distribution of the amplitudes of such jumps has  width of only
about ${H\over 2\pi}$. This means that we should live very close to the center
of the giant fluctuation, and the difference of energy densities between the
place where we live and the center of the ``bubble'' should be only about the
same amplitude as the typical perturbative fluctuation  $\delta\rho \sim
5\cdot10^{-5} \rho_c$. In other words, we should live very close to the  center
of the nearly perfect spherically symmetric bubble, which contains matter with
a smaller energy density than the matter outside it.

It is very tempting to interpret this effect in such a way that the Universe
around us becomes locally open.  The true
description of this effect is, of course, much more complicated; perhaps we
should see  the Hubble constant  decreasing at large distances.
This effect is extremely unusual. We became partially satisfied by our
understanding of this effect only after we confirmed its existence   by four
different methods, including computer simulations \cite{OPEN}. However, it may
happen that what we have found is simply a mathematical property of some
particular hypersurfaces in inflationary universe, and it does not have any
implications for the part of the Universe where we live.

Indeed, it is quite legitimate to use the distributions like $P_p$ for
descriptions of the structure of inflationary universe. However, it is not
quite clear whether one can use them to evaluate probabilities. For example,
instead of using the distribution
 $P_p(\phi,\chi,t)$ one may use the distribution $P_p(\phi,\chi \tau)$, where
$\tau \sim \log a$, and many of our result (though not all of them) will change
dramatically \cite{LLM,GBLL}. Still another answer will be obtained if one uses
some other cut-off procedure, see  \cite{VilNew}. The source of this
ambiguity can be easily understood. The total volume of all parts of an
eternally inflating Universe is infinite in the limit $t \to \infty$ (or $\tau
\to \infty$). Therefore when we are trying to compare volumes of domains with
different properties, we are comparing infinities. This leads to answers
depending on the way we are making this comparison.

It is possible that eventually we will resolve this problem. Still it will not
guarantee that we are on the right track.
Our use of $P_p$ as a probability measure was based on two hidden assumptions.
The first assumption is that   we are typical observers. The second assumption
is that the number of typical observers is directly proportional to the volume
of the Universe.
If this is correct, then we should live in the place where most observers live,
which should  correspond  to a maximum of $P_p$.

 However, is it absolutely clear that the probability for an observer to be
born in a particular part of the Universe is directly proportional to its
volume, or one should take into account something else? One cannot get any crop
even from a very large field without
having seeds first. The idea that life appears automatically
once there is enough space to be populated may be too primitive.
It is based on the assumption that one can describe emergence of life solely in
terms of physics.  It is
certainly a most economical approach, and one should try to go as far as
possible without invoking additional hypotheses. However, one should keep in
mind that this approach may  happen to be incomplete, especially if
consciousness has its own
degrees of freedom \cite{MyBook,Page}.

Another related question is whether we are actually typical? Does it make any
sense for each of us to calculate  {\it  a posteriori} what was the probability
to be born Russian, Italian or Chinese? Should we insist on our own
mediocrity, or,  {\it  vice versa}, should we try to explain why are we so
special? After all, for a long time we thought that we had the aristocratic
privilege to be the most intelligent species in the Universe. This, of course,
may be  wrong. Still, before using probabilities to calculate the likelihood of
our existence in a particular part of the Universe, it may be a good idea  to
learn more about ourselves.  I would take a certain risk to make a conjecture
that until we understand what is our life and what is the nature of
consciousness   our understanding of quantum cosmology will remain
fundamentally incomplete.

\subsection{A more radical approach: comparing Universes with different
coupling constants}
Previously we compared volume of the parts of the Universe with some particular
properties within one Universe. A more ambiguous program is based on a
combination of the baby Universe theory and stochastic approach to inflation.
The idea is that the coupling constants may take different values in different
Universes, or, more precisely, in different quantum states of the Universe
\cite{Coleman}. If this is the case, then perhaps we should live in those
Universes where conditions are better and the total volume suitable for life is
greater \cite{LinCosm}--\cite{GBL}.

The total volume is given by   ${\cal V}(\phi,\chi,t) = e^{ d(\lambda)  H_{\rm
\max}(\lambda)t}\, P_p(\phi,\chi,t)$. The first term in this expression is
especially important. If (and this is a big ``if''!) one can compare the
volumes of different Universes with different coupling constants at  the same
time $t$, the greatest volume will be occupied by the Universes with the
largest product $d(\lambda)  H_{\rm \max}(\lambda)$. For stationary
$P_p(\phi,\chi,t) = P_p(\phi,\chi)$ the exponential growth of ${\cal
V}(\phi,\chi,t)$ in the state with the largest $d(\lambda)  H_{\rm
\max}(\lambda)$ eventually beats all anthropic considerations. This may lead to
a very sharp prediction of the coupling constants  which maximize $d(\lambda)
H_{\rm \max}(\lambda)$.

Unfortunately,   this       immediately leads  to a trouble. For example, in
our investigation of the theory ${\lambda\over 4}\phi^4$ we have found that
$H_{\rm  max}= 2\sqrt{2\pi\over 3}$ does not depend on $\lambda$, whereas the
fractal dimension $d(\lambda)$ has its maximum $d = 3$ in the limit $\lambda =
0$. This is a rather general conclusion which seems to suggest that the
inflationary effective potential should be absolutely flat. But then there will
be no density perturbations which are necessary for galaxy formation. One may
try to avoid the problem with density perturbations assuming that they will be
produced by cosmic strings \cite{Vil,Al}, but in the theory with absolutely
flat potentials there will be no reheating and no cosmic strings. One may argue
that this means that the potential should be {\it almost} flat, i.e. that it
should be curved just enough   to allow baryons and strings to be produced and
life to appear. In fact, in such a case strings are not necessary. For example,
one may consider the hybrid inflation model   \cite{Hybrid}.  In this model
one can have good inflation and sufficiently large density perturbations
without any need for cosmic strings even if the potential is extremely (though
not exactly) flat. But the   problem is that the gain in $e^{ d(\lambda)
H_{\rm \max}(\lambda)t}$ eventually always beats the anthropic considerations,
which pushes us towards the models with {\it exactly} flat potentials. If the
effective potential is exactly flat, we have no reheating and no regular
density perturbations, but even in this case life may appear in an infinite
empty Universe with a very small but finite probability due to extremely
improbable quantum fluctuations. Even though  such conditions are extremely
improbable, eventually we will be compensated by the indefinitely large growth
of  volume due to the term $e^{ d(\lambda)  H_{\rm \max}(\lambda)t}$.  However,
in such a scenario there is no reason for our part of the Universe to be
homogeneous on the scale $10^{28}$ cm, which is much greater than what is
needed for our existence.
One may also argue that if quantum cosmology pushes us outside of  the limits
of our normal existence, it probably puts us at the verge of being immediately
extinct.

Another example is related to the cosmological constant problem. Adding it to
the Lagrangian also tends to increase $d(\lambda)$. Thus the considerations
based on the investigation of the factor $e^{ d(\lambda)  H_{\rm
\max}(\lambda)t}$ may push us towards very large values of the vacuum energy
density \cite{Vil}. Of course, one cannot go too far since our life cannot
exist if the vacuum energy density is too large. However, anthropic
considerations allow vacuum energy density $V_0$ two orders of magnitude
greater than the critical density $\sim 10^{-29}$g$\cdot$cm$^{-3}$, i.e.  two
orders of magnitude greater than
the present observational constraints on $V_0$  \cite{Weinberg82}. Moreover, as
we just mentioned,  the rapidly growing factor $e^{ d(\lambda)  H_{\rm
\max}(\lambda)t}$ should beat all  anthropic considerations and should push
$V_0$ even higher, which would be in a definite contradiction with the
observational data.

This indicates that   something should be modified either in the   radical
approach described in this subsection  or in our choice of the theories to
which we applied this approach \cite{GBL}. Each of these possibilities can be
true. First of all, it is not quite clear whether it makes any sense at all to
compare volume  of different Universes (rather than volume of different parts
of the same Universe) at the same time. Then, in certain theories the
probability distribution  ${\cal V}(\phi,\chi,t)$ is not stationary
\cite{GBLL}, so it cannot be represented as $e^{ d(\lambda)  H_{\rm
\max}(\lambda)t}\, P_p(\phi,\chi)$. Finally, under certain conditions the
fastest growth of ${\cal V}(\phi,\chi,t)$ appears in the theories where the
effective potential is not flat and the cosmological constant is not large
\cite{GBL}. For example, adding a positive cosmological constant  in the
Starobinsky model  {\it  decreases} the rate of  expansion. This pushes the
cosmological constant to zero \cite{GBL}. Unfortunately, this cannot be
considered as a possible solution of the cosmological constant problem since
the same mechanism may push the cosmological constant even further, toward its
negative values. To solve the cosmological constant problem it would be
necessary to find a mechanism which pushes it to zero from both sides.

It would be premature to make any final conclusions about the radical approach
described above. The  idea to use stochastic approach to inflation in order to
understand our place in the world is extremely attractive. However, this
powerful weapon should be used with caution, especially when one tries to
extend its limits of applicability and use it in the context of  the baby
Universe theory.
A possible attitude towards this approach is to consider it as a kind of
``theoretical experiment.''  We may try to use probabilistic considerations in
our trial-and-error approach to quantum cosmology. If we get unreasonable
results, this may serve as an indication that we are using quantum cosmology
incorrectly. However, if we   solve some problems  which could not be solved in
any other way, then we will have a reason to believe that we are moving in the
right direction.  In our opinion, at the present moment  we do not have
sufficient reasons to believe that the effective potential should be exactly
flat, that the density perturbations should be produced by strings appearing
after inflation, and that the cosmological constant should be as large as
possible. On the other hand, it is not excluded that the stochastic approach to
inflation,  or some of its generalizations,  will help us to solve the
cosmological constant problem. This possibility certainly deserves further
investigation.
We will return to the possibility of making predictions and calculating
probabilities in quantum cosmology in the next section, where we will consider
the model of an open inflationary universe.

\

\section{Inflation with $\Omega \not = 1$}
\subsection{Inflation and flatness of the Universe}
One of the most robust predictions of inflationary cosmology is that the
Universe after inflation becomes extremely flat, which corresponds to $\Omega =
1$. Here $\Omega = {\rho\over \rho_c}$,\,  $\rho_c$ being the energy density of
a
flat Universe.  There were many good reasons to believe that this prediction
was quite generic.  The only way to avoid this conclusion is to assume that the
Universe inflated only by about $e^{60}$ times. Exact value of the number of
e-foldings $N$ depends on details of the theory and may somewhat differ from
60. It is important, however, that in any particular theory  inflation by
extra 2 or 3 e-foldings would make  the Universe with $\Omega = 0.5$ or with
$\Omega = 1.5$ almost exactly flat. Meanwhile, the typical number of
e-foldings   in chaotic inflation scenario in the theory ${m^2\over 2}
\phi^2$ is not 60 but rather $10^{12}$.

One can construct models where
inflation leads to expansion of the Universe by the factor $e^{60}$. However,
in most of such models small number of e-foldings simultaneously implies that
  density perturbations
are extremely large.   It may be possible to overcome   this obstacle by a
specific choice of the
effective potential. However, this would be only a partial solution. If
the Universe
does not inflate long enough to become flat, then by the same token it
does not inflate long enough to become homogeneous and isotropic.
Thus,   the main reason why it is difficult to construct inflationary models
with $\Omega \not = 1$ is not the issue of fine tuning of the parameters of the
models, which is necessary to obtain the Universe inflating exactly $e^{60}$
times, but the problem of obtaining a homogeneous Universe after inflation.

Fortunately, it is possible to solve this problem, both for a closed Universe
 \cite{Lab} and for an open one \cite{Gott}--\cite{Arthur}. The
main idea is to use the well known fact that the region of space created in the
process of a quantum tunneling tends to have a spherically symmetric shape,
and homogeneous interior, if the tunneling process is suppressed strongly
enough. Then such bubbles of a new phase  tend  to evolve (expand) in a
spherically symmetric
fashion. Thus, if one
could associate the whole visible part of the Universe with an interior of one
such region, one would solve the homogeneity problem, and then all other
problems
will be solved by the subsequent relatively short stage of inflation.

For a closed Universe the realization of this program is relatively
straightforward
 \cite{Lab,Omega}. One should consider the process of quantum creation of  a
closed inflationary universe from ``nothing.''  If the probability of such a
process is exponentially suppressed (and this is indeed the case if inflation
is possible only at the energy density much smaller than the Planck density
 \cite{Creation}), then the Universe created that way will be  rather
homogeneous from the very beginning.

The situation with an open Universe is much more complicated. Indeed, an open
Universe is infinite, and it may seem impossible to create an infinite Universe
by a tunneling process. Fortunately, this is not the case: any bubble formed in
the process of the false vacuum decay looks from inside like an infinite open
Universe \cite{CL}.
 If this Universe continues inflating
inside the bubble \cite{Gott}--\cite{Arthur}, then we obtain an open
inflationary
Universe.

These possibilities became a subject of an active investigation only very
recently, and there are still many questions to be addressed. First of all, the
bubbles created by tunneling are not  {\it  absolutely} uniform even if the
probability of tunneling is very small.   This may easily spoil the whole
scenario since in the
end of the day we need to explain why the microwave background radiation is
isotropic with an accuracy of about $10^{-5}$. Previously we did not care much
about initial homogeneities, but if the stage of inflation is short, we will
the see original
inhomogeneities imprinted in the perturbations of the microwave background
radiation.

The second problem is to construct  realistic inflationary models where all
these ideas could be realized in a natural way. Whereas for the closed Universe
this problem can be easily solved \cite{Lab,Omega}, for an open Universe we
again meet complications. It would be very nice to to obtain an open Universe
in a theory of just one scalar  field \cite{BGT}. However, in practice it is
not very easy to obtain a satisfactory model of this type. Typically one is
forced either to introduce very complicated effective potentials, or consider
theories with nonminimal kinetic terms for the inflaton field \cite{Bucher}.
This makes the models   not only
fine-tuned, but also rather complicated. It  is very good to know that  the
models of such type in principle can be constructed, but  it is also very
tempting to find a
more natural realization of the   inflationary universe scenario which would
give
inflation with $\Omega < 1$.

 This goal
can be achieved if one considers models of two
scalar fields \cite{Omega}. One of them may be the standard inflaton
field
$\phi$ with a relatively small mass, another may be, e.g., the scalar field
responsible for the symmetry breaking in GUTs. The presence of two scalar
fields allows one to obtain the required bending of the inflaton potential by
simply changing the definition of the inflaton field in the process of
inflation. At the first stage the role of the inflaton is played by a heavy
field with a steep barrier in its potential, while on the second stage the
role of the inflaton is played by a light field, rolling in a flat direction
``orthogonal'' to the direction of quantum tunneling. This change of the
direction of evolution in the space of scalar fields removes the naturalness
constraints for the form of the potential, which are present in the case of one
field.

Inflationary models of this type
are quite simple, yet they have many interesting features. In these models
the Universe consists of infinitely many expanding bubbles immersed into
exponentially expanding false vacuum state. Each of these bubbles inside looks
like an open Universe, but the values of $\Omega$ in these Universes may take
any value from $1$ to $0$.
In some of these models the situation is even more complicated: Interior of
each bubble looks like an infinite  Universe with an effective value of
$\Omega$
slowly decreasing to $\Omega = 0$ at an exponentially  large distance from the
center  of the bubble. We will call such Universes quasiopen. Thus, rather
unexpectedly, we are obtaining a large  variety of  interesting and previously
unexplored possibilities. Our discussion of these possibilities will follow our
recent paper with Arthur Mezhlumian \cite{Arthur}.

\subsection{\label{Bubbles} Tunneling probability and spherical symmetry}

Typically it is assumed that the bubbles containing   open Universes are
exactly
spherically symmetric (or, to be more accurate, $O(3,1)$-symmetric \cite{CL}).
Meanwhile in realistic situations this condition may be violated for several
reasons. First of all, the bubble may be formed not quite symmetric. Then  its
shape may change even further due to growth of its initial  inhomogeneities and
due to quantum fluctuations which appear during the bubble wall expansion. As
we will see, this may cause a lot of problems if one wishes to maintain the
degree of anisotropy of the microwave background radiation inside the bubble at
the level of $10^{-5}$.

First of all, let us consider the issue of symmetry of a bubble at the moment
of its formation. For simplicity we will investigate the models where tunneling
can be described in the thin wall approximation. We will neglect gravitational
effects, which is possible as far as the initial radius $r$ of the bubble is
much smaller than $H^{-1}$. In this approximation (which works rather well for
the models to be discussed)
euclidean action of the $O(4)$-symmetric instanton describing bubble formation
is given by
\begin{equation}\label{o6}
S  = - {\epsilon\over 2} \pi^2 r^4 + 2\pi^2 r^3 s \  .
\end{equation}
Here $r$ is the radius of the bubble at the moment of its formation, $\epsilon$
is the difference of $V(\phi)$ between the false vacuum $\phi_{\rm initial}$
and the true vacuum $\phi_{\rm final}$, and $s$ is the surface tension,
\begin{equation}\label{o7}
s = \, \int_{\phi_{\rm initial}}^{\phi_{\rm final}} \sqrt{ 2(V(\phi) -
V(\phi_{\rm
final}))}\,  d\phi \  .
\end{equation}
The radius of the bubble can be obtained from the extremum of  (\ref{o6})
with respect to $r$:
\begin{equation}\label{o8}
r = {3s\over \epsilon } \ .
\end{equation}
Let us check how the action $S $ will change if one consider a bubble of a
radius $r + \Delta r$. Since the first derivative of $S $ at its extremum
vanishes, the change will be determined by its second derivative,
\begin{equation}\label{09}
\Delta S = {1\over 2} S'' (\Delta r)^2 = 9\pi^2\, {s^2\over \epsilon}\, (\Delta
r)^2 \ .
\end{equation}
 Now we should remember that all trajectories which have an action different
from the action at extremum by no more than  $1$ are quite legitimate. Thus the
typical deviation of the radius of the bubble from its classical value
(\ref{o8}) can be estimated from the condition $\Delta S \sim 1$, which gives
\begin{equation}\label{o10}
|\Delta r| \sim {\sqrt\epsilon\over 3\pi \,s} \ .
\end{equation}
Note, that even though we considered spherically symmetric perturbations, our
estimate  is based on corrections proportional to $(\delta r)^2$, and
therefore it should remain valid for perturbations which have an amplitude
$\Delta r$, but change their sign in different parts of the bubble surface.
Thus, eq. (\ref{o10}) gives an estimate of a typical degree of asymmetry of the
bubble at the moment of its creation:
\begin{equation}\label{o11}
A(r) \equiv  {|\Delta r| \over r} \sim {\epsilon\sqrt\epsilon\over 3\pi \,s^2}
\ .
\end{equation}
This simple estimate exactly coincides with the corresponding result obtained
by Garriga and Vilenkin \cite{VilGarr} in their study of quantum fluctuations
of bubble walls. It was shown in \cite{VilGarr} that when an   empty  bubble
begins
expanding, the typical deviation $\Delta r$ remains constant. Therefore the
asymmetry given by the ratio ${|\Delta r| \over r}$ gradually vanishes. This is
a pretty general result: Waves produced by a brick falling to a pond do not
have the shape of a brick, but gradually become circles.

However, in our case the situation is somewhat more complicated. The wavefront
produced by a brick in inflationary background preserves the shape of the brick
if its size  is much greater than $H^{-1}$. Indeed, the wavefront moves   with
the speed approaching the speed of light, whereas the distance between
different parts of a region with initial size greater than $H^{-1}$ grows with
a much greater (and ever increasing) speed. This means that inflation
stretches the wavefront without changing its shape on scale much greater than
$H^{-1}$. Therefore during inflation which
continues inside the bubble the symmetrization of its shape   occurs only in
the very beginning, until the radius of the bubble approaches $H^{-1}$. At this
first stage expansion of the bubble occurs mainly due to the motion of the
walls rather than due to inflationary stretching of the Universe, and our
estimate of the bubble wall asymmetry as well as the results obtained by
Garriga and Vilenkin for the empty bubble remain valid. At the moment when the
radius of the bubble becomes equal to $H^{-1}$ its asymmetry
becomes
\begin{equation}\label{o12}
A(H^{-1}) \sim {|\Delta r| H} \sim {\sqrt\epsilon H\over 3\pi \,s} \ ,
\end{equation}
and the subsequent expansion of the bubble does not change this value very
much. Note that the Hubble constant here is determined by the vacuum energy
{\it
after} the tunneling, which may differ from the initial energy density
$\epsilon$.

The deviation of the shape of the
bubble from spherical symmetry implies that the beginning of the second stage
of inflation inside the bubble will be not exactly synchronous, with the delay
time $\Delta t \sim \Delta r$. This, as usual,  may lead to adiabatic density
perturbations on the
horizon scale of the order of $H\Delta t$, which coincides with the bubble
asymmetry $A$ after its size becomes greater than $H^{-1}$, see Eq.\
(\ref{o12}).

To estimate this contribution to density perturbations, let us consider again
the
simplest model with the effective potential
\begin{equation}\label{o1}
V(\phi) = {m^2\over 2} \phi^2 - {\delta\over 3} \phi^3 + {\lambda\over 4}\phi^4
\ .
\end{equation}
Now we will consider it
in the limit  where the two minima of this potential have almost
the same depth, which is necessary for  validity of the thin wall
approximation. In this case    $2\delta^2 = 9 M^2\lambda$, and the effective
potential (\ref{o1}) looks approximately like ${\lambda \over 4} \phi^2 (\phi
- \phi_0)^2$, where $\phi_0 = {2\delta\over 3\lambda} = \sqrt {2\over \lambda}
{M}$ is the position of the local minimum of the effective
potential.  The surface tension in this model is given by $s =
\sqrt{\lambda\over 2} {\phi_0^3\over 6} = { M^3\over 3\lambda}$ \cite{Tunn}.
We will also introduce a phenomenological parameter $\mu$, such that $\mu
{M^4\over 16\lambda} = \epsilon$. The smallness of this parameter controls
applicability of the thin-wall approximation, since the value of the effective
potential near the top of the potential barrier   at $\phi = \phi_0/2$ is given
by $M^4\over 16\lambda$.   Then our estimate of density
perturbations associated with the bubble wall (\ref{o12})    gives
\begin{equation}\label{o14}
\left. {\delta\rho\over \rho}\right|_{\rm bubble}  \sim A(H^{-1}) \sim
{\sqrt{\mu\lambda}
H\over  4 \pi
M}   \ .
\end{equation}
Here $H$ is the value of the Hubble constant at the beginning of inflation
inside the bubble.

In order to have  $\left. {\delta\rho\over \rho}\right|_{\rm bubble} {\
\lower-1.2pt\vbox{\hbox{\rlap{$<$}\lower5pt\vbox{\hbox{$\sim$}}}}\ } 5 \times
10^{-5}$ (the number $5 \times 10^{-5}$ corresponds to the amplitude of density
perturbations in  the COBE normalization) one
should have
\begin{equation}\label{o17}
\left. {\delta\rho\over \rho}\right|_{\rm bubble}  \sim \, {\sqrt{\mu\lambda}
H\over  4 \pi
M}\,  {\ \lower-1.2pt\vbox{\hbox{\rlap{$<$}\lower5pt\vbox{\hbox{$\sim$}}}}\ }\,
 5 \times 10^{-5} \ .
\end{equation}
For $H\ll M$ perturbations produced by the bubble walls   may   be sufficiently
small  even if the coupling constants are relatively large and the bubbles at
the moment of their formation are very inhomogeneous.

There is a long way from our simple estimates  to the full theory of
anisotropies of
cosmic microwave background induced by fluctuations of the domain wall. In
particular, the significance of this effect will clearly depend on the value of
$\Omega$  \cite{Open}.
 The constraint (\ref{o17}) may appear  only if one can ``see'' the
scale at which the bubble walls have imprinted their fluctuations. If inflation
is long enough,
this scale becomes
exponentially large, we do not see the fluctuations due to bubble walls, but
then we return to the standard
inflationary scenario of a flat inflationary universe. However, for $\Omega \ll
1$ inflation is short, and it does not preclude us from seeing perturbations in
a vicinity of the bubble walls. In such a case one should   take
the constraint  (\ref{o17}) very seriously.

\subsection{\label{Simplest} The simplest model of a (quasi)open inflationary
Universe}

 In this section we will explore an extremely  simple model of two
scalar fields, where the Universe after inflation becomes open (or quasiopen,
see below) in a very natural way \cite{Omega}.

Consider a model of
two noninteracting scalar fields, $\phi$ and $\sigma$, with the effective
potential
\begin{equation}\label{3}
V(\phi, \sigma) = {m^2\over 2}\phi^2 + V(\sigma) \ .
\end{equation}
Here $\phi$ is a weakly interacting inflaton field, and $\sigma$, for example,
can be the field responsible for the symmetry breaking in GUTs. We will assume
that $V(\sigma)$ has a local minimum at $\sigma = 0$, and a global minimum at
$\sigma_0 \not = 0$, just as in the old inflationary
theory. For definiteness, we will assume that this potential is given by
${M^2\over 2} \sigma^2 -
{\alpha M } \sigma^3 + {\lambda\over 4}\sigma^4 + V(0)$, with $V(0) \sim
{M^4\over 4 \lambda}$, but it is not essential;
no fine tuning of the shape of this potential will be required.

Note that so far we did not make any unreasonable complications to the standard
chaotic inflation scenario; at large $\phi$ inflation is driven
by the field $\phi$, and the GUT potential is necessary in the theory anyway.
In order to obtain density perturbations of the necessary amplitude the mass
$m$ of the scalar field $\phi$ should be of the order of $10^{-6} M_{\rm P}
\sim
10^{13}$ GeV \cite{MyBook}.

Inflation begins at $V(\phi, \sigma) \sim M_{\rm P}^4$. At this stage
fluctuations of
both fields are very strong, and the Universe enters the stage of
self-reproduction, which finishes for the field $\phi$ only when it becomes
smaller than $M_{\rm P} \sqrt{M_{\rm P}\over m}$ and the energy density drops
down to $m
M_{\rm P}^3  \sim 10^{-6} M_{\rm P}^4$ \cite{MyBook}. Quantum fluctuations of
the field
$\sigma$ in some parts of the Universe put it directly to the absolute minimum
of
$V(\sigma)$, but in some other parts the scalar field $\sigma$ appears in the
local minimum of $V(\sigma)$ at $\sigma  = 0$. We will follow evolution of such
domains. Since the energy density in such
domains will be greater, their volume will  grow  with a greater speed, and
therefore they will be especially important for us.

One may worry that all
domains with $\sigma = 0$
will tunnel to the minimum of $V(\sigma)$ at the stage when the field $\phi$
was very large and quantum fluctuations of the both fields were large too.
This may happen if the Hubble constant induced by the scalar field $\phi$ is
much greater than the curvature of the potential $V(\sigma)$:
\begin{equation}\label{s1}
{m\phi\over M_{\rm P}} {\
\lower-1.2pt\vbox{\hbox{\rlap{$>$}\lower5pt\vbox{\hbox{$\sim$}}}}\ } M \ .
\end{equation}

This decay can be easily suppressed if one introduces a
small interaction $g^2\phi^2\sigma^2$ between these two fields, which
stabilizes the state with $\sigma = 0$ at large $\phi$. Another possibility is
to add a
nonminimal interaction with gravity of the form $-{\xi\over 2} R\phi^2$, which
makes inflation impossible for $\phi > {M_{\rm P}\over 8\phi\xi}$. In this case
the
condition (\ref{s1}) will never be satisfied.  However, there is a much simpler
answer to this worry. If the effective potential of the field $\phi$ is so
large that the field $\sigma$ can easily jump to the true minimum of
$V(\sigma)$,
then the Universe becomes divided into infinitely many domains with all
possible values of $\sigma$ distributed in the following way
 \cite{Star,MyBook}:
\begin{equation}\label{s2}
{P(\sigma= 0)\over P(\sigma = \sigma_0)} \sim \exp\left({3M^4_{\rm P}\over 8
V(\phi,0)} - {3M^4_{\rm P}\over 8V(\phi,\sigma)}\right) = \exp\left({3M^4_{\rm
P}\over 4(m^2\phi^2 + 2V(0))} - {3M^4_{\rm P}\over 4 m^2\phi^2}\right)\ .
\end{equation}
One can easily check that at the moment when the field $\phi$ decreases to ${M
M_{\rm P}\over m}$ and  the condition (\ref{s1}) becomes violated, we will
have
\begin{equation}\label{s3}
{P(0)\over P(\sigma_0)}  \sim \exp\left(-{C\over \lambda}\right) \ ,
\end{equation}
where $C$ is some constant, $C = O(1)$. After this moment the probability of
the false vacuum decay typically becomes much smaller. Thus the fraction of
space which survives in the false vacuum state $\sigma = 0$ until this time
typically is very small, but finite (and calculable). It is important,   that
these rare domains with $\sigma = 0$ eventually will dominate the volume of the
Universe since if the probability of the false vacuum decay is small enough,
the volume of the domains in the false vacuum will continue growing
exponentially without end.

The main idea of our scenario can be explained as follows. Because the fields
$\sigma$ and
$\phi$ do not interact with each other, and the dependence of the probability
of tunneling on the vacuum energy at the GUT scale is negligibly small
 \cite{CL}, tunneling to the minimum of $V(\sigma)$ may occur with
approximately
equal
probability at all sufficiently small values of the field $\phi$ (see, however,
below). The
parameters of the bubbles of the field $\sigma$ are determined by the mass
scale $M$ corresponding to the effective potential $V(\sigma)$. This mass scale
in
our model is much greater than $m$. Thus the duration of tunneling in the
Euclidean ``time'' is much smaller than $m^{-1}$. Therefore the field $\phi$
practically does not change its value during the tunneling.  If
the probability of decay at a given $\phi$ is small enough, then it does not
destroy the whole vacuum state $\sigma = 0$ \cite{GW}; the bubbles of the new
phase are produced all the way when    the field $\phi$ rolls down to $\phi =
0$. In this process  the Universe  becomes filled with
(nonoverlapping) bubbles immersed in the false vacuum state with $\sigma = 0$.
Interior of each of these bubbles   represents an open Universe. However, these
bubbles   contain  {\it  different} values of the field $\phi$, depending on
the
value of this field at the  moment when the bubble formation occurred. If the
field $\phi$ inside a bubble is smaller than $3 M_{\rm P}$, then the Universe
inside
this bubble will have a vanishingly small $\Omega$, at the age $10^{10}$ years
after the end of inflation it will be practically empty, and life of our type
would not exist there.  If the field $\phi$ is much greater than $3 M_{\rm P}$,
the
Universe inside the bubble will be almost exactly flat, $\Omega = 1$, as in the
simplest version of the chaotic inflation scenario. It is important, however,
that  {\it  in  an eternally existing self-reproducing Universe there will be
infinitely many Universes containing any particular value of $\Omega$, from
$\Omega = 0$ to $\Omega = 1$}, and one does not need any fine tuning of the
effective potential to obtain a Universe with, say,  $0.2 <\Omega < 0.3$

Of course, one can argue that we did not solve the problem of fine tuning, we
just transformed it into the fact that only a very small percentage of all
Universes will have  $0.2 <\Omega < 0.3$. However, first of all, we
achieved our goal in a very simple theory, which does not require any
artificial potential bending and nonminimal kinetic terms. Then, there may be
some reasons why it is preferable for us to live in a Universe with a small
(but not vanishingly small) $\Omega$.

The simplest way to approach this problem is to find  how the probability
for the bubble production depends on $\phi$. As we already pointed out, for
small $\phi$ this dependence is not very strong. On the other hand, at large
$\phi$ the probability rapidly grows and  becomes quite large at $\phi > {M
M_{\rm P}\over m}$. This may suggest that the bubble production typically
occurs at
$\phi > {M M_{\rm P}\over m}$, and then for ${M\over  m} \gg 3$ we typically
obtain
flat Universes, $\Omega = 1$. This is another manifestation of the problem of
premature decay of the state $\sigma = 0$ which we discussed above. Moreover,
even if the probability to produce the Universes with different $\phi$ were
entirely $\phi$-independent, one could argue that the main volume of the
habitable parts of the Universe is contained in the bubbles with $\Omega = 1$,
since  the interior of each such bubble  inflated longer. Indeed, the total
volume of each bubble created in a state with  the field $\phi$ during
inflation in our model grows by the factor of $\exp{6\pi\phi^2\over M_{\rm
P}^2}$ \cite{MyBook}. It seems clear that the bubbles with greater $\phi$ will
give the largest contribution to the total volume of the Universe after
inflation. This would be the simplest argument in favor of the standard
prediction $\Omega = 1$ even in our   class of models.

However, there exist
several ways of resolving this problem: involving coupling $g^2\phi^2\sigma^2$,
which stabilizes the state $\sigma = 0$ at large $\phi$, or adding nonminimal
interaction with gravity of the form $-{\xi\over 2} R\phi^2$. In either way one
can
easily suppress production of the Universes with   $\Omega = 1$. Then the
maximum of probability will correspond to some value $\Omega < 1$, which can be
made equal to any given number from $1$ to $0$ by changing the parameters $g^2$
and $\xi$.

For example, let us   add  to the Lagrangian the term $-{\xi\over 2} R\phi^2$.
This term makes inflation impossible for $\phi > \phi_c = {M_{\rm P}\over
\sqrt{8\pi\xi}}$. If initial value of the field $\phi$ is much smaller than
$\phi_c$, the size of the Universe during inflation grows $\exp{2\pi\phi^2\over
M_{\rm P}^2}$ times, and the volume grows $\exp{6\pi\phi^2\over M_{\rm P}^2}$
times, as in the theory ${m^2\over 2} \phi^2$ with $\xi = 0$. For initial
$\phi$ approaching $\phi_c$ these expressions somewhat change, but in order to
get a very rough estimate of the increase of the size of the Universe in this
model (which is sufficient to get an illustration of our main idea) one can
still use the old expression $\exp{2\pi\phi^2\over M_{\rm P}^2}$. This
expression reaches its maximum near $\phi = \phi_c$, at which point the
effective gravitational constant becomes infinitely large and inflationary
regime ceases to exist \cite{Maeda,GBL}. Thus, one may argue that in this case
the main part of the volume of the Universe will appear from the bubbles with
initial value of the field $\phi$ close to $\phi_c$. For $\xi \ll 4.4\times
10^{-3}$ one has $\phi_c \gg 3 M_{\rm P}$. In this case  one would have typical
Universes expanding much more than $e^{60}$ times, and therefore $\Omega
\approx 1$. For    $\xi \gg 4.4\times 10^{-3}$ one has $\phi_c \ll 3 M_{\rm
P}$, and therefore  one would have  $\Omega \ll 1$ in all inflationary bubbles.
It is clear that by choosing particular values of the constant  $\xi$ in the
range of  $\xi \sim 4.4\times 10^{-3}$ one can obtain the distribution of the
Universes with the maximum of the distribution concentrated near any desirable
value of $\Omega < 1$.
Note that the position of the peak of the distribution is very sensitive to the
value of $\xi$: to have the peak concentrated in the region $0.2 < \Omega <
0.3$ one would have to fix $\xi$ (i.e. $\phi_c$) with an accuracy of few
percent. Thus, in this approach to the calculation of probabilities to live in
a Universe with a given value of $\Omega$ we still have the problem of fine
tuning.

However, calculation of probabilities in the context of the theory of a
self-reproducing Universe is a very ambiguous process, and it is even not quite
clear that this process makes any sense at all. For example, we may
formulate the problem in a different way. Consider a domain of the false vacuum
with $\sigma = 0$ and $\phi = \phi_1$. After some evolution it
produces one or many bubbles with $\sigma = \sigma_0$ and the field $\phi$
which after some time becomes equal to $\phi_2$. One may argue that the most
efficient way this process may go is the way which in the end produces the
greater volume. Indeed, for the inhabitants of a bubble it does not matter how
much time did it take for this process to occur. The total number of
observers produced by this process will depend on the total volume of the
Universe at the hypersurface of a given density, i.e. on the hypersurface of a
given $\phi$. If the domain instantaneously  tunnels to the state $\sigma_0$
and $\phi_1$, and then the field $\phi$ in this domain slowly rolls from
$\phi_1$ to $\phi_2$, then the volume of this domain grows $\exp
\Bigl({2\pi\over M_{\rm P}^2} (\phi_1^2 -\phi_2^2)\Bigr)$ times \cite{MyBook}.
Meanwhile, if the tunneling takes a long time, then the field $\phi$ rolls down
extremely slowly being in the false vacuum state with $\sigma = 0$. In this
state the Universe expands much faster than in the state with $\sigma =
\sigma_0$. Since it expands much faster, and it takes the field much longer to
roll from $\phi_1$ to $\phi_2$, the trajectories of this kind bring us much
greater volume. This may serve as an argument that most of the volume is
produced by the bubbles created at a very small $\phi$, which leads to the
Universes with very small $\Omega$.

One may use another set of considerations, studying all trajectories beginning
at $\phi_1, t_1$ and ending at $\phi_2, t_2$. This will bring us  another
answer, or, to be more precise, another set of answers, which will depend on
the choice of the time parametrization \cite{LLM}.  Still another answer will
be obtained by the method
  recently proposed by Vilenkin, who suggested to introduce a particular
cutoff procedure which (almost) completely eliminates dependence of the final
answer on the time parametrization \cite{VilNew}. A more radical possibility
would be to integrate over all time parametrizations. This task is very
complicated, but it   would completely eliminate dependence of the final answer
on the time parametrization.

There is a very deep  reason why the calculation of the probability to obtain a
Universe with a given $\Omega$ is so ambiguous. We have discussed this reason
in Sect. 3.1 in general terms; let us see how the situation looks in
application to the open Universe scenario. For those who  lives inside
a bubble there is be no way to say at which stage (at which
time from the point of view of an external observer) this bubble was produced.
Therefore one should compare  {\it  all} of these bubbles produced at all
possible times.  The self-reproducing Universe should exist for indefinitely
long time, and therefore   it should contain  infinitely many bubbles with all
possible values of $\Omega$. Comparing infinities is a very ambiguous task,
which gives results depending on the procedure of comparison. For example, one
can consider an infinitely large box of apples and an infinitely large box of
oranges. One may pick up one apple and one orange, then one apple and one
orange, over and over again, and conclude that there is an equal number of
apples and oranges. However, one may also pick up one apple and
two oranges, and then one apple and two oranges again, and conclude that there
is twice as many oranges as apples. The same situation happens when one tries
to compare the number of  bubbles with different values of $\Omega$. If we
would know how to
solve  the problem of measure in quantum cosmology, perhaps we would be able to
obtain
something similar to an open Universe in the trivial $\lambda\phi^4$ theory
without any first order phase transitions \cite{OPEN}, see Sect. 3.1.  In the
meantime, it is already encouraging that in our scenario
there are infinitely many inflationary universes with all possible value of
$\Omega < 1$. We can hardly live in the empty bubbles with $\Omega = 0$. As for
the choice between the bubbles with different nonvanishing values of $\Omega <
1$,  it is quite possible that eventually we will find out an unambiguous way
of predicting the most probable value of $\Omega$, and we are going to continue
our work in this direction. However,  as we already discussed in the previous
section, it might also happen that this question
is as meaningless as the question whether it is more probable to be born as a
Chinese rather than as an Italian. It is quite conceivable that the only way to
find out in which of the bubbles do we   live   is to make observations.

Some words of caution are in order here. The bubbles produced in our simple
model
are not  {\it  exactly} open Universes. Indeed, in the models discussed in
 \cite{CL}--\cite{BGT} the time of reheating (and the temperature of the
Universe after the reheating) was synchronized with the value of the scalar
field inside the bubble. In our case the situation is very similar, but not
exactly. Suppose that the Hubble constant induced by $V(0)$ is much
greater than the Hubble constant related to the energy density of the scalar
field $\phi$. Then the speed of rolling of the scalar field $\phi$ sharply
increases inside the bubble. Thus, in our case the field $\sigma$ synchronizes
the motion of the field $\phi$, and then the hypersurface of a constant field
$\phi$ determines the hypersurface of a constant temperature. In the models
where the rolling of the field $\phi$ can occur only inside the bubble (we will
discuss such a model shortly)  the  synchronization is precise, and everything
goes as in the models of refs. \cite{CL}--\cite{BGT}. However, in our simple
model the scalar field $\phi$ moves down outside the bubble as well, even
though it does it very slowly. Thus,  synchronization of   motion of the
fields $\sigma$ and $\phi$  is not precise; hypersurface of a constant $\sigma$
ceases to be a hypersurface of a constant density. For example, suppose that
the field $\phi$ has taken some value $\phi_0$ near the bubble wall when the
bubble was just formed. Then the bubble expands, and during this time the field
$\phi$ outside the wall  decreases, as $\exp \Bigl(-{m^2t\over 3 H_1}\Bigr)$,
where $H_1 \approx  H(\phi = \sigma = 0)$ is the Hubble constant at the first
stage of inflation, $H_1 \approx \sqrt{8\pi V(0)\over 3 M_{\rm P}^2}$. At the
moment
when the bubble expands $e^{60}$ times, the field $\phi$ in the region just
reached by  the bubble wall decreases to  $\phi_o\exp \Bigl(-{20 m^2\over
H^2_1}\Bigr)$ from its original value $\phi_0$. the Universe inside the bubble
is a homogeneous open Universe only if this change is negligibly small. This
may not be a real problem. Indeed,  let us assume that $V(0) ={\tilde M}^4$,
where ${\tilde M} =
10^{17}$ GeV. (Typically the energy density scale $\tilde M$ is related to the
particle mass as follows: ${\tilde M} \sim \lambda^{-1/4} M$.) In this case
$H_1 = 1.7 \times 10^{15}$ GeV, and for $m =
10^{13}$ GeV one obtains ${20 m^2\over   H_1^2} \sim 10^{-4}$. In such a case
a typical degree of distortion of the picture of a homogeneous open Universe is
very small.

Still this issue requires careful investigation. When the bubble wall continues
expanding even further, the scalar field outside of it eventually drops down to
zero. Then there will be no new matter created near the wall.  Instead of
infinitely large homogeneous open Universes we are obtaining   spherically
symmetric islands of a size much greater than the size of the observable part
of our Universe. We do not know whether this unusual picture is an advantage or
a
disadvantage of our model. Is it possible to consider different parts of the
same
exponentially large island as domains of different ``effective'' $\Omega$? Can
we attribute some part of the dipole anisotropy of the microwave background
radiation to the possibility that we live somewhere outside of the center of
such island? In any
case, as we already mentioned, in the limit $m^2 \ll H_1^2$
we do not expect that the small deviations of the geometry of space inside the
bubble from the geometry of an open Universe can do much harm to our model.

Our model admits many generalizations, and details of the scenario which we
just discussed depend on the values of parameters. Let us forget for a moment
about all complicated processes which occur  when the field $\phi$ is rolling
down to $\phi = 0$, since this part of the picture depends on the validity of
our ideas about initial conditions. For example, there may be no
self-reproduction of inflationary domains with large $\phi$ if one considers an
effective
potential of the field $\phi$ which is very curved at large $\phi$. However,
there will be self-reproduction of the Universe in a state
$\phi = \sigma = 0$, as in the old inflation scenario. Then the main portion of
the volume of the Universe will be determined by the processes which occur when
   the fields    $\phi$  and $\sigma$    stay  at the local minimum of the
effective potential, $\phi = \sigma = 0$.   For definiteness we will assume
here that $V(0) = {\tilde M}^4$, where ${\tilde M}$ is   the   stringy scale,
${\tilde M} \sim 10^{17} -
10^{18}$ GeV. Then the Hubble constant $H_1 = \sqrt{8\pi V(0)\over 3M^2_{\rm
P}} \sim
\sqrt{8\pi \over 3} {{\tilde M}^2\over M_{\rm P}}$ created by the energy
density
$V(0)$ is
much greater than $m \sim 10^{13}$ GeV. In such a case the scalar field $\phi$
will not stay exactly at $\phi = 0$. It will  be  relatively homogeneous on the
horizon scale $H_1^{-1}$, but otherwise it will  be chaotically distributed
with
the dispersion $\langle\phi^2\rangle = {3H^4\over 8\pi^2m^2}$ \cite{MyBook}.
This means that the field $\phi$ inside each of the bubbles produced by the
decay of the false vacuum can take any value $\phi$ with the probability
\begin{equation}\label{4}
P \sim \exp\left(-{\phi^2\over 2 \langle\phi^2\rangle}\right) \sim
\exp\left(-{3m^2 \phi^2M_{\rm P}^4\over 16 {\tilde M}^8}\right) \ .
\end{equation}
One can check that for ${\tilde M} \sim 4.3\times10^{17}$ GeV the typical value
of the
field $\phi$ inside the bubbles will be $\sim 3\times 10^{19}$ GeV. Thus, for
${\tilde M} > 4.3\times10^{17}$ GeV most of the Universes produced during the
vacuum
decay will be flat, for ${\tilde M} < 4.3\times10^{17}$ GeV most of them will
be open.
It is interesting that in this version of our model the percentage of open
Universes is determined by the stringy scale (or by the GUT scale). However,
since the process of bubble production in this scenario   goes without   end,
the total  number of Universes with any particular value of  $\Omega < 1$ will
be infinitely large   for any value of ${\tilde M}$.   Thus this   model shows
us is the
simplest way to resurrect some of the ideas of the old inflationary theory with
the help of chaotic inflation, and simultaneously to obtain  inflationary
Universe with $\Omega < 1$.

Note that this version of our model will not suffer for the problem of
incomplete synchronization. Indeed, the average value of the field $\phi$ in
the false vacuum outside the bubble will remain constant until the bubble
triggers its decrease.
However, this model, just as its previous version, may suffer from another
problem. The Hubble constant $H_1$ before the tunneling in this model was much
greater
than the Hubble constant $H_2$ at the beginning of the second stage of
inflation. Therefore the fluctuations of the
scalar field before the tunneling were very large, $\delta \phi \sim {H_1\over
2 \pi}$, much greater than the
fluctuations generated after the tunneling,  $\delta \phi \sim {H_2\over 2
\pi}$. This may lead to very large
density perturbations on the scale comparable to the size of the bubble. For
the models with $\Omega = 1$ this effect would not cause any problems since
such
perturbations would be far away over the present particle horizon, but for
small $\Omega$  this
may lead to unacceptable anisotropy of the microwave background radiation.

Fortunately, this may not be a real difficulty. A possible solution is very
similar to the bubble symmetrization described in the previous section.

Indeed, let us consider more carefully how the long wave perturbations produced
outside the bubble may penetrate into it. At the moment when the bubble is
formed, it has a   size (\ref{o8}), which is  smaller than $H_1^{-1}$
 \cite{CL}. Then the bubble walls begin moving with the speed gradually
approaching the speed of light. At this stage the comoving size of the bubble
(from the point of view of the original coordinate system in the false vacuum)
grows like
\begin{equation}\label{n1}
r(t) = \int_{0}^{t}{dt e^{-H_1 t}} = H_1^{-1} (1 - e^{-H_1 t}) \ .
\end{equation}
During this time the fluctuations of the scalar field $\phi$ of the amplitude
${H_1\over 2\pi}$ and of the wavelength $H_1^{-1}$, which previously were
outside the bubble, gradually become covered by it. When these perturbations
are outside the bubble, inflation with the Hubble constant $H_1$ prevents them
from oscillating and moving. However, once these perturbations penetrate inside
the bubble, their amplitude becomes decreasing \cite{MZ,SP}. Indeed, since the
wavelength of the perturbations is $\sim H_1^{-1} \ll H_2^{-1} \ll m^{-1}$,
these
perturbations move inside the bubbles as relativistic particles, their
wavelength grow  as $a(t)$, and their amplitude decreases just like an
amplitude of electromagnetic field, $\delta\phi \sim a^{-1}(t)$, where $a$ is
the scale factor of the Universe inside a bubble \cite{MZ}. This process
continues until the wavelength of each perturbation reaches $H_2^{-1}$ (already
at the second stage of inflation). During this time the wavelength grows
${H_1\over H_2}$ times, and the amplitude decreases ${H_2\over H_1}$ times, to
become the standard amplitude of perturbations produced at the second stage of
inflation: $  {H_2\over H_1}\, {H_1\over 2\pi} = {H_2\over 2\pi}$.

In fact, one may  argue that this computation was too naive, and that these
perturbations should be neglected altogether. Typically we treat long wave
perturbations in inflationary universe like classical wave for the reason that
the waves with the wavelength much greater than the horizon can be interpreted
as states with extremely large occupation numbers \cite{MyBook}. However, when
the new  born perturbations (i.e. fluctuations which did not acquire an
exponentially large wavelength yet) enter  the bubble (i.e. under the horizon),
they effectively return to the realm of quantum fluctuations again. Then one
may argue that one should simply forget about the waves with the wavelengths
small enough to fit into the bubble, and consider perturbations created at the
second stage of inflation not as a result of stretching of these waves, but as
a new process of creation of perturbations of an amplitude ${H_2\over 2\pi}$.

One may worry   that perturbations which had wavelengths somewhat greater than
$H_1^{-1}$ at the moment of the bubble formation  cannot completely penetrate
into the bubble. If, for example, the field $\phi$ differs from some constant
by $+{H_1\over 2\pi}$ at the distance $H_1^{-1}$ to the left   of the bubble at
the moment of its formation, and by  $-{H_1\over 2\pi}$ at the distance
$H_1^{-1}$ to the  right of the bubble, then this difference remains frozen
independently of all processes inside the bubble. This may suggest that there
is some unavoidable asymmetry of the distribution of the field inside the
bubble. However, the field inside the bubble will not be distributed like a
straight line slowly rising from  $-{H_1\over 2\pi}$ to  $+{H_1\over 2\pi}$.
Inside
the bubble the field will be almost homogeneous; the inhomogeneity $\delta \phi
\sim -{H_1\over 2\pi}$ will be concentrated only in a small vicinity near the
bubble wall.

Finally we should verify that this scenario  leads to bubbles which are
symmetric enough, see eq. (\ref{o17}). Fortunately, here we do not have any
problems. One can easily check that for our model with $m \sim 10^{13}$ GeV and
$\tilde M \sim \lambda^{-1/4} M > 10^{17} GeV$ the condition (\ref{o17}) can be
  satisfied even for not very small values of the coupling constant $\lambda$.

The arguments presented above should be confirmed by a more detailed
investigation of the vacuum structure inside the expanding bubble in our
scenario. If, as we hope,  the result of the investigation will be positive, we
will have an
extremely simple model of an open inflationary universe. In the meantime, it
would be nice to have a model where we do not have any problems at all with
synchronization and
with  large fluctuations on the scalar field in the false vacuum. We will
consider such a model in the next section.

\subsection{\label{Hybrid} Hybrid inflation  and  natural inflation with
$\Omega < 1$}

 The model to be discussed below is a version of the hybrid
inflation scenario \cite{Hybrid}, which is a slight generalization (and a
simplification) of our previous model
(\ref{3}):
\begin{equation}\label{4a}
V(\phi,\sigma) = {g^2\over 2}\phi^2\sigma^2 + V(\sigma) \ .
\end{equation}
We eliminated the massive term of the field $\phi$ and added explicitly the
interaction ${g^2\over 2}\phi^2\sigma^2$, which, as we have mentioned already,
can be useful (though not necessary)  for stabilization of the state $\sigma =
0$ at large $\phi$. Note
that in this model the line $\sigma = 0$ is a flat direction in the
($\phi,\sigma$) plane. At large $\phi$ the only minimum of the effective
potential with respect to $\sigma$ is at the line $\sigma = 0$.  To give a
particular example, one can take $V(\sigma) = {M^2\over 2} \sigma^2 -{\alpha M
} \sigma^3 + {\lambda\over 4}\sigma^4 +V_0$. Here $V_0$ is a constant which is
added to ensure that $V(\phi,\sigma) = 0$ at the absolute minimum of
$V(\phi,\sigma)$.  In this case the minimum of the potential $V(\phi,\sigma)$
at $\sigma \not = 0$ is deeper than the minimum at $\sigma = 0$ only for $\phi
< \phi_c$, where $\phi_c = {M\over g}\sqrt{{2\alpha^2\over  \lambda} -1}$. This
minimum for $\phi = \phi_c$ appears at $\sigma = \sigma_c = {2\alpha M\over
\lambda}$.

The bubble formation becomes possible only for $\phi < \phi_c$. After the
tunneling the field $\phi$ acquires an effective mass $m = g\sigma$ and begins
to move towards $\phi = 0$, which provides the mechanism for the second stage
of inflation inside the bubble. In this scenario evolution of the scalar field
$\phi$ is exactly synchronized with the evolution of the field $\sigma$, and
the Universe inside the bubble appears to be open.

Effective mass of the   field $\phi$ at the minimum of $V(\phi,\sigma)$ with
$\phi = \phi_c$, $\sigma = \sigma_c = {2\alpha M\over  \lambda}$ is   $m =
g\sigma_c = {2g\alpha M\over  \lambda}$. With a decrease of the field $\phi$
its effective mass at the minimum of $V(\phi,\sigma)$ will grow, but not
significantly. For simplicity, we will consider the case $\lambda = \alpha^2$.
 In this case it can be shown that $V(0) = 2.77\, {M^4\over \lambda}$, and the
Hubble constant before the phase transition is given by $4.8\, {M^2\over \sqrt
\lambda M_{\rm P}}$.  One should check  what is necessary to avoid too large
density perturbations (\ref{o17}). However, one should take into account that
the mass $M$ in (\ref{o17}) corresponds to the  curvature of the effective
potential near $\phi = \phi_c$ rather than at $\phi = 0$. In our case this
implies that one should use $\sqrt 2 M$ instead of $M$ in this equation.  Then
one obtains the following constraint on the mass $M$: \ $M\sqrt \mu   {\
\lower-1.2pt\vbox{\hbox{\rlap{$<$}\lower5pt\vbox{\hbox{$\sim$}}}}\ }
2\times10^{15}$ GeV.  Note that   the thin wall approximation (requiring $\mu
\ll 1$) breaks down  far away from $\phi = \phi_c$. Therefore in general eq.
(\ref{o17}) should be somewhat improved.  However for $\phi \approx \phi_c$ it
works quite well.  To be on a safe side, we will take   $M = 5\times 10^{14}$
GeV.   Other parameters may vary; one may consider, e.g., the theory with  $g
\sim 10^{-5}$, which gives
 $\phi_c = {M\over g}  \sim 5\times 10^{19}\  \mbox{GeV} \sim 4M_{\rm P}$.
The effective mass $m$ after the phase transition is equal to ${2gM\over \sqrt
\lambda}$ at $\phi = \phi_c$, and then it grows by only $25\%$ when the field
$\phi$ changes all the way down from  $\phi_c$ to $\phi = 0$.   As we already
mentioned, in order to obtain the proper amplitude of density perturbations
produced by inflation
inside the bubble one should have $m \sim 10^{13}$ GeV. This corresponds to
$\lambda = \alpha^2 = 10^{-6}$.

The bubble
formation becomes possible only for $\phi < \phi_c$. If it happens in the
interval $4M_{\rm P} > \phi > 3 M_{\rm P}$, we obtain a flat Universe. If it
happens at $\phi < 3M_{\rm P}$, we obtain an open Universe. Depending on the
initial value of the field $\phi$, we can obtain all possible values of
$\Omega$, from $\Omega = 1$ to $\Omega = 0$. The value of the Hubble constant
at the minimum with $\sigma \not = 0$ at $\phi = 3M_{\rm P}$ in our model does
not differ much from the value of the Hubble constant before the bubble
formation. Therefore we do not expect any specific problems with the large
scale density perturbations in this model.
 Note also that the probability of tunneling at large $\phi$ is very small
since the depth of the minimum at $\phi \sim \phi_c$, $\sigma \sim \sigma_c$
does not differ much from the depth of the minimum at $\sigma = 0$, and there
is no tunneling at all for $\phi > \phi_c$. Therefore
the number of flat Universes produced by this mechanism will be strongly
suppressed as compared with the number of open Universes, the degree of this
suppression being very sensitive to the value of $\phi_c$. Meanwhile, life of
our type is impossible in empty Universes with $\Omega \ll 1$. This may provide
us with a tentative explanation of the small value of $\Omega$ in the context
of our model.

Another model  of inflation with $\Omega < 1$ is the based on a certain
modification of the ``natural inflation'' scenario  \cite{Natural}.  The main
idea is to take the effective potential of the ``natural inflation'' model,
which looks like a tilted Mexican hat,  and make a deep hole it its center at
$\phi = 0$  \cite{Arthur}. In the beginning inflation occurs near $\phi = 0$,
but then the bubbles with $\phi \not = 0$ appear. Depending on the phase of the
complex scalar field $\phi$ inside the bubble, the next stage of inflation,
which occurs just as in the old version of the ``natural inflation'' scenario,
leads to formation of the Universes with all possible values of $\Omega$.

A detailed discussion of this scenario can be found in  \cite{Arthur}; we will
not repeat it here. What is most important for us is that there exist several
rather simple  models of an open inflationary universe. Inflationary models
with $\Omega = 1$ admittedly are somewhat simpler. Therefore we still hope that
several years later we
will know that our Universe is flat, which will be a strong experimental
evidence in favor of  inflationary cosmology in its simplest form. However, if
observational data will show,
beyond any reasonable doubt, that $\Omega \not = 1$, it will not imply  that
inflationary theory is wrong. Indeed, now we know
that there is a large class of  internally consistent cosmological models which
may describe
creation of   large homogeneous Universes with all possible values of $\Omega$,
and  so far all of these models are based on inflationary cosmology.

\

\section { Reheating after inflation}

The theory of reheating of the Universe after inflation is the
most important application of the quantum theory of
particle creation, since almost all matter constituting the
Universe at the subsequent radiation-dominated stage was
created during this process \cite{MyBook}. At the stage of
inflation all
energy was
concentrated in a classical slowly moving inflaton field $\phi$. Soon
after the
end of inflation this field began to oscillate near the minimum of
its
effective potential. Gradually it produced many elementary particles,
they
interacted with each other and came to a state of thermal equilibrium
with some
temperature $T_r$, which was called the reheating temperature.

An elementary theory  of reheating was first
developed in   \cite{DolgLinde} for the new inflationary scenario.
Independently a theory of reheating in the $R^2$ inflation was constructed
in~\cite{st81}.  Various
aspects of this
theory   were further elaborated by many authors, see e.g.
\cite{Dolg,Brand}.
 Still, a general scenario of reheating was
absent. In particular, reheating in the chaotic inflation theory
remained
almost unexplored.
The present section contains results obtained recently in our work with Kofman
and Starobinsky \cite{KLSREH}.  We have found that the process of
reheating typically
consists of  three different stages.  At the first stage, which cannot be
described by the elementary theory of reheating,  the classical coherently
oscillating
inflaton field $\phi$ decays into massive bosons (in particular, into
$\phi$-particles) due to parametric resonance.  In many models the resonance is
very broad, and the process occurs extremely
rapidly (explosively).  Because of the Pauli exclusion principle, there is no
explosive  creation of fermions.
To distinguish this stage from the stage of  particle decay and thermalization,
we will call it {\it pre-heating}. Bosons produced at that stage are far away
from thermal equilibrium and typically have enormously large occupation
numbers. The second stage  is the decay of previously produced  particles. This
stage typically can be described by methods developed in   \cite{DolgLinde}.
However,
these methods should be applied not  to the decay of the original homogeneous
inflaton field, but to the decay of particles and fields produced at the  stage
of explosive reheating. This considerably changes many features of  the
process, including the final value of the reheating temperature.  The
third stage is the stage of thermalization, which can be described by
standard
methods, see e.g. \cite{MyBook,DolgLinde}; we will not consider it here.
Sometimes this stage
may occur simultaneously with the second one. In our
investigation we
have used the formalism of the
 time-dependent Bogoliubov transformations   to find the density of
created particles, $n_{\vec k}(t)$.
A detailed
description of this theory will be given in \cite{REH}; here we will
outline
our main conclusions using a simple semiclassical
approach.

  We will consider a simple chaotic inflation scenario describing the
classical inflaton scalar field
$\phi$ with the effective potential   $V(\phi) =  \pm {1\over2}
m_\phi^2 \phi^2+{\lambda\over 4}\phi^4$. Minus sign corresponds to
spontaneous symmetry breaking $\phi \to \phi +\sigma$ with generation of a
classical scalar field $\sigma = {m_\phi \over\sqrt\lambda}$.  The field $\phi$
after inflation may decay
into bosons $\chi$ and fermions $\psi$ due to the interaction  terms $- {
1\over2} g^2 \phi^2 \chi^2$ and
 $- h \bar \psi \psi \phi$. Here $\lambda$, $ g$ and  $h$ are
small coupling constants.  In case of  spontaneous symmetry breaking, the term
$- {
1\over2} g^2 \phi^2 \chi^2$ gives rise to the   term  $- g^2 \sigma\phi
\chi^2$. We will assume for simplicity that the bare masses
of the fields $\chi$ and $\psi$ are very small, so that one can write  $ m_\chi
(\phi) =
  g \phi$,  $m_{\psi}(\phi) =  |h\phi|$.

Let us briefly recall the elementary theory of reheating
\cite{MyBook}.  At
$\phi > M_{\rm P}$, we have a stage of inflation.  This stage is supported
by the
friction-like term $3H\dot\phi$ in the equation of motion for the scalar
field. Here $H\equiv \dot a/a$ is the Hubble parameter, $a(t)$ is the
scale factor of the Universe.
However, with a decrease of the field $\phi$ this term becomes less
and less important, and inflation ends at $\phi {\
\lower-1.2pt\vbox{\hbox{\rlap{$<$}\lower5pt\vbox{\hbox{$\sim$}}}}\ }M_{\rm
P}/2$.
After that the
field $\phi$  begins  oscillating near the minimum of
$V(\phi)$. The amplitude of the oscillations  gradually
decreases because of expansion of the
Universe, and also because of the energy transfer to particles
created by the
oscillating field.  Elementary
theory of reheating is based on the
 assumption that  the
classical oscillating scalar field $\phi (t)$ can be represented as a
collection of scalar particles at rest. Then the rate of decrease of
the energy of oscillations
coincides with the decay rate  of  $\phi$-particles. The
rates of
the processes $\phi \to \chi\chi$  and $\phi \to  \psi\psi$ (for  $m_\phi \gg
2m_\chi, 2m_\psi$) are given
by
 \begin{equation}\label{7}
  \Gamma ( \phi \to \chi \chi) =  { g^4 \sigma^2\over 8
\pi m_{\phi}}\  , \ \ \ \ \
\Gamma( \phi \to \psi \psi )  =  { h^2 m_{\phi}\over 8 \pi}\ .
 \end{equation}
Reheating
completes when the rate of expansion of the Universe given   by the Hubble
constant $H=\sqrt{8\pi \rho\over 3 M^2_{\rm P}} \sim  t^{-1}$ becomes smaller
than
the total decay rate $\Gamma =  \Gamma (\phi \to \chi \chi) + \Gamma
(\phi \to
\psi \psi )$. The reheating temperature can be estimated by
$T_r \simeq 0.1\, \sqrt{\Gamma M_{\rm P}}$\,.

It is interesting to note that in accordance with the elementary theory of
reheating the amplitude squared of the oscillating scalar field decays
exponentially, as $e^{-\Gamma t}$. Phenomenologically, this can be described by
adding the term $\Gamma\dot\phi$ to the equation of motion of the scalar field.
Unfortunately, many authors took this prescription too seriously and
investigated the possibility that the term $\Gamma\dot\phi$, just like the term
$3H\dot\phi$, can support inflation. We should emphasize \cite{KLSREH}, that
adding the term $\Gamma\dot\phi$ to the equation of motion is justified only at
the stage of oscillations (i.e. after the end of inflation), and only for the
description of the {\it amplitude of oscillations} of the scalar field, rather
than for the description of the scalar field itself.  Moreover, even at the
stage of oscillations this description becomes incorrect as soon as the
resonance effects become important.

As we already mentioned, elementary theory of reheating can provide a
qualitatively correct
description of particle decay at the last stages of reheating.
Moreover, this theory  is always applicable  if the inflaton field
can decay  into fermions only, with a small coupling constant $h^2 \ll
m_{\phi}/M_{\rm P}$.
 However,
typically this theory is inapplicable to the description of the first stages of
reheating, which makes the whole process quite different. In what follows we
will develop the theory of the first stages of reheating. We will begin with
the theory of a massive scalar field $\phi$ decaying into particles $\chi$,
then we consider the theory  ${\lambda\over 4} \phi^4$,
and finally we will discuss reheating in the theories with spontaneous symmetry
breaking.

We begin with the investigation of the simplest  inflationary
model with the effective potential
${m^2_\phi\over 2}\phi^2$.
Suppose that this field  only interacts
with a light  scalar field $\chi$
 ($m_{\chi} \ll m_{\phi}$) due to the
 term $-{ 1\over2} g^2 \phi^2  \chi^2$.
The equation for quantum fluctuations of the field $\chi$
with the physical momentum $\vec k/a(t)$ has the following form:
\begin{equation}\label{M}
\ddot \chi_k   + 3H \dot \chi_k +  \left({k^2\over a^2(t)}
+ g^2 \Phi^2\, \sin^2(m_{\phi}t) \right) \chi_k = 0 \ ,
\end{equation}
where $k = \sqrt {\vec k^2}$, and $\Phi$ stands for the amplitude of
oscillations of the field $\phi$. As we shall see, the
main contribution to $\chi$-particle
production is given by excitations of the field $\chi$ with
 $k/a \gg m_\phi$, which is much
greater than $H$ at the stage
of oscillations. Therefore, in the first approximation we may neglect
the expansion of the Universe,  taking $a(t)$ as a constant and omitting
the term $3H \dot \chi_k$ in (\ref{M}). Then the equation (\ref{M})
describes an oscillator with a variable
frequency $\Omega_k^2(t)=
 k^2a^{-2} + g^2\Phi^2\, \sin^2(m_{\phi}t) $.
Particle production occurs due to a
nonadiabatic change of this frequency. Equation (\ref{M}) can be
reduced to the well-known  Mathieu equation:
\begin{equation}\label{M1}
\chi_k''   +   \left(A(k) - 2q \cos 2z \right) \chi_k = 0 \ ,
\end{equation}
where   $A(k)
= {k^2 \over m_\phi^2 a^2}+2q$, $q = {g^2\Phi^2\over
4m_\phi^2} $, $z
= m_{\phi}t$, prime denotes differentiation with respect to $z$.
An important property of solutions of the equation (\ref{M1}) is the
existence of an exponential instability $\chi_k \propto \exp
(\mu_k^{(n)}z)$ within the set of resonance bands  of frequencies
$\Delta k^{(n)}$ labeled by an integer index $n$.
This instability corresponds to exponential growth of occupation
numbers of quantum fluctuations
$n_{\vec k}(t) \propto \exp (2\mu_k^{(n)} m_{\phi} t)$
  that may be interpreted as particle
production.   As one
can show,   near the line $A = 2q$ there are regions in the first,
the second and the  higher instability bands
where the unstable modes grow extremely
rapidly, with $\mu_k \sim 0.2$. We will show analytically in~\cite{REH} that
for $q \gg 1$
 typically
 $\mu_k \sim {\ln 3\over 2\pi}
\approx 0.175$ in the instability bands along the line $A = 2q$,
but its maximal value is ${\ln(1+\sqrt{2}) \over \pi} \approx 0.28$.
Creation of
particles in the regime of a broad  resonance ($q > 1$) with $2\pi \mu_k =
O(1)$ is very different from that in  the usually
considered case of a narrow resonance ($ q \ll 1$),
 where $2\pi \mu_k \ll 1$.
 In particular, it
proceeds during a tiny part of each oscillation of the field $\phi$
when $1-\cos z \sim q^{-1}$ and the induced effective mass of the
field $\chi$ (which is
determined by the condition $m^2_{\chi}= g^2\Phi^2/2$) is less than
$m_{\phi}$.
 As a result, the number of
particles grows exponentially within just a few oscillations of the
field
$\phi$. This leads to an extremely rapid  (explosive)  decay of the
classical
scalar field $\phi$.
This regime occurs only
if  $q {\ \lower-1.2pt\vbox{\hbox{\rlap{$>$}\lower5pt\vbox{\hbox{$\sim$}}}}\ }
\pi^{-1}$, i.e. for $g\Phi {\
\lower-1.2pt\vbox{\hbox{\rlap{$>$}\lower5pt\vbox{\hbox{$\sim$}}}}\ }
m_\phi$, so that $m_\phi \ll gM_{\rm P}$ is the necessary condition for it.
One can show that a typical energy $E$ of a particle produced at this stage is
determined by
 equation $A-2q \sim \sqrt{q}$, and is given by
 $E  \sim  \sqrt{g m_\phi M_{\rm P}}$ \cite{REH}.

Creation of $\chi$-particles leads to the two main effects:
transfer of the energy from the homogeneous field $\phi (t)$ to these
particles and generation of the contribution to the effective mass of
the $\phi$ field:  $m^2_{\phi ,eff}=m^2_{\phi}+g^2\langle\chi^2
\rangle_{ren}$.
The last term in the latter expression
quickly becomes larger than
$m^2_{\phi}$. One should take
both these effects into account when calculating backreaction of
created particles on the process.
As a result, the stage of the broad resonance creation ends up within
the short time
$t\sim m_{\phi}^{-1} \ln (m_{\phi}/g^5M_{\rm P})$,
when $\Phi^2 \sim
\langle\chi^2\rangle$ and  $q = {g^2\Phi^2\over
4m_{\phi ,eff}^2}$
becomes smaller than $1$.
At this time the energy density of produced particles
$\sim E^2 \langle\chi^2\rangle \sim g m_\phi M_{\rm P} \Phi^2$ is of the same
order as the original energy density
$\sim {m_\phi^2} M_{\rm P}^2$ of the scalar field
$\phi$ at the end of inflation. This gives the amplitude of
oscillations at the end of the stage of the broad resonance particle creation:
$\Phi^2 \sim \langle\chi^2\rangle \sim
g^{-1} m_\phi M_{\rm P} \ll M_{\rm P}^2$.
Since $E\gg m_{\phi}$, the effective equation of state of the whole
system becomes $p\approx \rho /3$. Thus, explosive creation
practically eliminates a prolonged intermediate matter-dominated stage
after the end of inflation which was thought to be characteristic
to many inflationary models.
However, this does not mean that the process of reheating has been completed.
Instead of $\chi$-particles in the thermal equilibrium with
 a typical energy
 $E \sim T \sim (mM_{\rm P})^{1/2}$, one has particles with a much
smaller energy $\sim  (g m_\phi M_{\rm P})^{1/2}$,
but with extremely large
mean occupation numbers  $n_k \sim g^{-2} \gg 1$.

After that the Universe expands as $a(t)\propto \sqrt t$, and
the scalar field $\phi$ continues its decay in the regime of the narrow
resonance creation $q\approx {\Phi^2\over 4 \langle\chi^2\rangle}
\ll 1$. As a result,
$\phi$ decreases rather slowly, $\phi \propto t^{-3/4}$.
This regime is very important  because  it makes the energy of the
$\phi$ field much smaller than that of the $\chi$-particles.
One can show that the decay finally stops when the amplitude of
oscillations $\Phi$ becomes smaller than $g^{-1} m_\phi$ \cite{REH}.
This happens at the moment $t\sim  m_{\phi}^{-1}  (gM_{\rm P}/m_{\phi})^{1/3}$
(in the case   $m < g^7 M_{\rm P}$ decay ends somewhat later,
in the perturbative regime).
The physical reason why the decay  stops is rather general: decay of
the particles $\phi$ in our model occurs due to its interaction with another
$\phi$-particle (interaction term is quadratic in $\phi$ and in $\chi$). When
the  field $\phi$ (or the number of $\phi$-particles) becomes
small, this process
is inefficient.  The scalar field can decay completely only if a
single  scalar $\phi$-particle can decay into other  particles, due
to the processes
$\phi \to \chi \chi$ or $\phi \to \psi \psi$, see eq. (\ref{7}). If
 there is
 no spontaneous symmetry breaking and no interactions with fermions
in our model, such  processes are impossible.

At later stages the energy of oscillations of the inflaton field
decreases as $a^{-3}(t)$, i.e. more slowly than the decrease of
energy of hot ultrarelativistic matter $\propto a^{-4}(t)$. Therefore, the
relative contribution of the field $\phi(t)$ to the total energy density
of the Universe
rapidly grows. This   gives rise to an unexpected possibility that
the inflaton field  by
itself, or other scalar fields can be
cold dark matter candidates, {\it even if they strongly interact with each
other}. However, this possibility requires
a certain degree of fine tuning; a more immediate application of our result is
that it allows one to rule out a wide class of inflationary models which do not
contain interaction terms of the type of $g^2\sigma\phi\chi^2$ or
$h\phi\bar\psi\psi$.

 So far we have not considered  the term ${\lambda \over 4} \phi^4$
in the effective potential. Meanwhile this term leads to production
of $\phi$-particles, which in some cases appears to be the leading
effect.
 Let us  study the $\phi$-particle production in the theory
with $V(\phi)  =
{m^2_{\phi}\over 2} \phi^2 + {\lambda\over 4}\phi^4$ with $m^2_{\phi}
\ll \lambda  M_{\rm P}^2$. In this case the effective potential
of the field $\phi$ soon after the end of inflation at
$\phi \sim M_{\rm P}$ is dominated by the term
${\lambda\over 4} \phi^4$.  Oscillations  of the field $\phi$ in this
theory
are not sinusoidal, they are
given by elliptic functions, but with a good accuracy one can write
$\phi(t)
\sim \Phi \sin (c\sqrt \lambda \int \Phi dt)$, where
$c={\Gamma^2(3/4)\over \sqrt \pi} \approx 0.85$.   the Universe at
that time expands as at the
radiation-dominated stage: $a(t)\propto \sqrt t$. If one neglects
the feedback of created $\phi$-particles on the homogeneous field
$\phi (t)$, then its amplitude $\Phi (t) \propto a^{-1}(t)$, so that $a\Phi
=const$.
Using a conformal time $\eta$, exact equation for quantum fluctuations
 $\delta \phi$
 of the field $\phi$ can be reduced to the Lame equation. The results remain
essentially the same if we use an  approximate equation
\begin{equation}\label{lam1}
{d^2(\delta\phi_k)\over d\eta^2}   +   {\Bigl[{k^2} +
3\lambda a^2\Phi^2\, \sin^2 (c\sqrt\lambda a\Phi \eta)\Bigr]}
\delta\phi_k = 0,~~\eta =\int {dt\over a(t)}={2t\over a(t)}\, ,
\end{equation}
which leads to the Mathieu equation with $A =
{k^2\over c^2\lambda a^2\Phi^2} +
{3\over 2c^2} \approx  {k^2\over c^2\lambda a^2\Phi^2} + 2.08$, and
$q = {3\over 4c^2} \approx 1.04$.   Looking at the instability chart, we see
that the
resonance occurs in the second band, for $k^2 \sim 3\lambda a^2\Phi^2$. The
maximal value of the coefficient $\mu_k$ in   this band for $q \sim 1$
approximately equals to $0.07$. As long as the backreaction of created
particles is small, expansion of the Universe does not shift fluctuations away
from the resonance band, and the
number of produced particles grows  as  $\exp (2c\mu_k\sqrt\lambda a\Phi \eta)
\sim
\exp ({\sqrt\lambda\Phi t\over 5})$.

After the time interval $\sim M_{\rm P}^{-1}\lambda^{-1/2}|\ln \lambda|$,
  backreaction of created particles  becomes significant. The growth of the
fluctuations
$\langle\phi^2\rangle$ gives rise to a   contribution
$3\lambda \langle\phi^2\rangle$ to the effective mass squared of the field
$\phi$, both in the equation for $\phi (t)$ and in Eq. (\ref{lam1}) for
inhomogeneous modes.
The stage of explosive reheating ends when $\langle\phi^2\rangle$ becomes
greater than $\Phi^2$. After that, $\Phi^2 \ll
\langle\phi^2\rangle$ and
the effective frequency of oscillations is determined by the
term $\sqrt{3\lambda \langle\phi^2\rangle}$.
The corresponding process is
 described by Eq. (\ref{M1}) with  $A(k) = 1 + 2q + {k^2
\over
3\lambda a^2\langle\phi^2\rangle}$, $q = {\Phi^2\over 4
\langle\phi^2\rangle}$. In this regime $q \ll 1$, and particle creation occurs
in the  narrow resonance regime in the second band with $A \approx 4$.  Decay
of the field in this regime is extremely slow: one can show \cite{REH} that the
amplitude $\Phi$ decreases
only by a factor
 $t^{1/12}$ faster that  it would decrease without any decay, due to the
expansion of the Universe only, i.e., $\Phi \propto t^{-7/12}$.
Reheating stops altogether when the presence of non-zero mass
$m_{\phi}$ though still small as compared to $\sqrt{3\lambda
\langle\phi^2\rangle}$
appears enough for the expansion of the Universe to drive
a mode away
from the narrow resonance. It happens when the amplitude $\Phi$ drops
up to a value $\sim m_{\phi}/\sqrt \lambda$.

In addition to this process, the field $\phi$ may decay  to
$\chi$-particles.
This is the leading process for    $g^2\gg \lambda$.
The equation for $\chi_k$ quanta has  the same form as eq.
(\ref{lam1})
with the obvious change $\lambda \to g^2/3$.
Initially  parametric resonance is broad. The values of the parameter
$\mu_k$
along the line $A = 2q$ do not change monotonically, but typically
for $q \gg
1$ they are 3 to 4 times greater than the parameter $\mu_k$ for the
decay of
the field $\phi$ into its own quanta. Therefore, this pre-heating
process is very
efficient. It ends at the moment $t\sim M_{\rm P}^{-1}\lambda^{-1/2}
\ln (\lambda /g^{10})$ when $\Phi^2 \sim \langle \chi^2
\rangle \sim g^{-1}\sqrt \lambda M_{\rm P}^2$. The typical energy of created
$\chi$-particles is $E \sim (g^2\lambda)^{1/4}M_{\rm P}$. The following
evolution is essentially the same as that described above for the case of a
massive scalar field decaying into $\chi$-particles.

Finally, let us consider the case with symmetry breaking.  In the
beginning, when the amplitude of oscillations is much greater than
$\sigma$, the theory of  decay of the inflaton field is the same as in the case
considered above. The most important part of  pre-heating occurs at this stage.
When the amplitude of the oscillations becomes smaller than
$m_\phi/\sqrt\lambda$ and the field begins oscillating near the minimum of the
effective potential at $\phi = \sigma$, particle production due to
the narrow parametric
resonance typically becomes  very weak.
The main reason for this is related to the backreaction of
particles created at the
preceding stage of pre-heating on the rate of expansion of the Universe and on
 the shape of the effective potential \cite{REH}. A rather interesting effect
which makes investigation of this regime especially complicated is a temporary
(non-thermal) symmetry restoration which occurs because of the interaction of
the field $\phi$ with its fluctuations $<\phi^2>$. Importance of
spontaneous symmetry
breaking for the theory of reheating should not be underestimated, since it
gives rise to the interaction term   $g^2\sigma\phi\chi^2$ which is linear in
$\phi$. Such terms are necessary for a complete decay of the inflaton field in
accordance with the perturbation theory (\ref{7}).

In this section we presented the new theory of reheating developed in
\cite{KLSREH}, where we performed an investigation of reheating with an account
of expansion of the Universe and of the  backreaction of created particles,
both in the broad resonance regime and in the narrow resonance case. As a
result of this investigation, we obtained equations for the power-law decrease
of the amplitude of an oscillating scalar field with an account taken of all of
these effects.
During the last  year   there appeared many other papers on the theory of
reheating \cite{Shtanov}--\cite{Kaiser}, which made the physical picture of
reheating  even more clear.  Unfortunately, it is not easy to compare the
results obtained in \cite{Shtanov}--\cite{Kaiser} with the results of our work
\cite{KLSREH}.  For example, a very thorough investigation of reheating in the
narrow resonance regime  without a complete account of backreaction was
performed in \cite{Shtanov,Yoshimura,Kaiser}, and their results in this
approximation agree with the corresponding results of \cite{KLSREH}. However,
as we have seen, at the first, most efficient stages of reheating   the
resonance is broad, and when it becomes narrow a complete account of
backreaction becomes necessary \cite{KLSREH}. Backreaction was studied in a
very detailed way in ref.~\cite{Boyan}, but their investigation was performed
neglecting expansion of the Universe, which was an important  part of our work.
 That is why in this review we  concentrated on the results obtained in
\cite{KLSREH}.
However, to obtain a complete theory of reheating a much more detailed
investigation will be necessary, and in this respect many of the results
obtained in \cite{Shtanov}--\cite{Kaiser} should be very useful.

We should emphasize   that the stage of parametric resonance is just the first
stage of the process. If one naively takes the energy density at the end of
explosive reheating and assumes that this energy density instantaneously
transfers to heat, one may overestimate the reheating temperature by many
orders of magnitude. Indeed, after the stage of explosive reheating the
bose-particles created at this stage have enormously large occupation numbers,
and they should further decay into the usual elementary particles. This may
take a lot of time, during which the energy density of the Universe may
decrease dramatically. To find the reheating temperature   one should
investigate the subsequent  decay of the  particles created at the stage of
explosive reheating. This decay can be described  by the old perturbative
methods developed in~\cite{DolgLinde}.
Note, however, that now
this  theory  should be applied not to the decay of the
original large and homogeneous oscillating inflaton field, but to the decay  of
particles produced at the stage of pre-heating, as well as to the decay of
small remnants of the  classical   inflaton field. This makes a lot
of difference, since typically coupling constants of interaction of the
inflaton field with matter are extremely small, whereas coupling constants
involved in the decay of  other bosons  can be much greater. As a result,
the reheating temperature can be much higher than the typical temperature $T_r
{\ \lower-1.2pt\vbox{\hbox{\rlap{$<$}\lower5pt\vbox{\hbox{$\sim$}}}}\ } 10^9$
GeV
which could be obtained
neglecting the stage of parametric resonance \cite{REH}. In addition, one
should make a careful study of the process of establishing of thermal
equilibrium \cite{Boyan2}.
On the other hand, such processes as baryon creation after inflation occur best
of all outside  the state of thermal equilibrium. Therefore, the stage
of   explosive reheating (pre-heating), which produces fields and particles
outside of the state of thermal equilibrium,  may play an extremely important
role in the cosmological
theory.  Another  consequence of the resonance effects is an almost
instantaneous change of equation of state from the vacuum-like one to the
equation of state of relativistic matter $p = \rho/3$.  This may be important
for   investigation of the primordial black hole formation, which could appear
from growing   density perturbations if equation of state after inflation for a
long time was $p = 0$.

A rather nontrivial example of reheating appears in  inflationary models based
on supergravity, see e.g. \cite{Nan}--\cite{Bert}.   The leading  mode of the
single-inflaton decay in such models often involves creation of  a gravitino,
which is a fermion.   This does not necessarily mean   that the   first
explosive stage cannot be realized in such models. Indeed, just as in the
theory ${\lambda\over 4}\phi^4$, at the first stage the homogeneous classical
oscillating  inflaton field $\phi$ may decay  into decoherent waves or
particles  of the same field $\phi$. However, this will be just a first stage
of reheating, after which one should consider decay of the inflaton particles
by the usual perturbative methods. In such a situation one does not expect any
deviations of the reheating temperature from its value obtained by perturbative
methods  \cite{DolgLinde}, \cite{Nan}--\cite{Bert}.

One should note also that in certain models the oscillations of the scalar
field from the very beginning occur in the region where the conditions for the
explosive reheating formulated in \cite{KLSREH} are not satisfied. Such a
situation occurs, e.g.,  in ``natural  inflation''  \cite{DolgF}, where the
change of the effective mass of the inflaton field during its oscillations is
relatively small, and the  conditions   of existence of narrow resonance in
expanding Universe derived in \cite{KLSREH} are violated.

Let us briefly summarize our results:

1. In many models where decay of the inflaton field can occur in the purely
bosonic sector the first stages of reheating occur due to parametric resonance.
This process (pre-heating) is extremely efficient even if the corresponding
coupling constants are very small.
However, there is no explosive reheating in the models where decay of the
inflaton field is necessarily accompanied by fermion production.

 2. The stage of explosive reheating due to a broad resonance typically is very
short. Later the resonance becomes narrow, and
finally the stage of pre-heating finishes altogether. Interactions of
particles produced at this stage, their decay into other particles and
subsequent thermalization typically
require  much more time that the stage of pre-heating, since these processes
are suppressed by the small
values of coupling constants.

3. The last stages of reheating typically
  can be described by the elementary theory of reheating \cite{DolgLinde}.
However, this theory should be applied not to the original inflaton field, but
to the products of its decay formed at the   stage of explosive reheating. In
some models it changes the final value of the reheating temperature.

4. Existence of the intermediate stage between the end of explosive reheating
and the beginning of  thermal equilibrium may have important implications for
the theory of baryogenesis.

5. Reheating never completes in the theories where a single $\phi$-particle
cannot decay into other particles. This implies that reheating completes only
if the theory  contains interaction terms like $\phi\sigma\chi^2$ of
$\phi\bar\psi\psi$. In most cases the theories where reheating never completes
contradict observational data.  On the other hand, this result suggests an
interesting possibility that  the classical scalar fields  (maybe even the
inflaton field itself) may be responsible for the
dark matter of the Universe even if they strongly interact with other matter
fields.

\section{Conclusions}

Inflationary theory is already  more than 15 years old, and its main principles
seem to be well understood. Nevertheless, it is young enough to bring us many
new surprises. Originally we expected that inflation was a short intermediate
stage after the hot big bang. Now it seems that the standard   big bang theory
is only a part of inflationary cosmology which describes local (but not global)
properties of the self-reproducing inflationary universe.  Even though each
part of the Universe   expands (or collapses), the Universe as a whole may be
stationary. One of the main purposes of inflationary cosmology was to solve the
primordial monopole problem by expanding the distance between the monopoles.
Recently we learned that the monopoles themselves may expand exponentially and
become as large as a universe \cite{LL}. On the other hand, we learned that an
infinitely large open inflationary universe may fit into an interior of a
single bubble of a finite size produced during the false vacuum decay. This
demonstrated that even though $\Omega = 1$ remains one of the rather robust
predictions of inflationary cosmology, it will be impossible to kill inflation
by proving that our universe is open. The process of creation of matter after
inflation also happened to be extremely interesting and complicated, involving
investigation of nonperturbative resonance effects in an expanding universe.
Rapid development of the inflationary theory is a very good sign indicating
that we are moving fast towards a complete cosmological theory -- assuming, as
we all hope, that   we have chosen the right direction.

\end{document}